\documentclass{aa}

\usepackage{gensymb}
\usepackage{upgreek}
\usepackage{color}

\usepackage[version=4]{mhchem}

\usepackage{hhline}
\usepackage{longtable}
\usepackage{graphicx}

\usepackage[varg]{txfonts}

\usepackage{upgreek} 
\usepackage{multirow} 
\usepackage[FIGTOPCAP, center, nooneline]{subfigure}
\usepackage[]{natbib}
\usepackage{upgreek}

\usepackage[varg]{txfonts}
\usepackage{microtype}
\usepackage{hyperref}
\usepackage{booktabs}
\usepackage{xcolor}
\usepackage{hyperref}
\usepackage{graphicx}
\usepackage{multirow}
\usepackage{float}
\usepackage[switch, modulo]{lineno}
\modulolinenumbers[1]
\linenumbers
\hypersetup{
  colorlinks=true,
  linkcolor=blue,
  citecolor=blue
}

\usepackage{etoolbox}

\let\oldcitet=\citet

\renewcommand{\citet}[1]{\textcolor[rgb]{0,0,1}{\oldcitet{#1}}}

\newcommand{\HII}{H\,{\sc ii}}

\nolinenumbers

\begin{document}

\title{Sub-arcsecond  imaging of HCN and CN in the Orion Bar}
\subtitle{UV-driven warm Nitrogen chemistry in dense environments}

\titlerunning{Sub-arcsecond  imaging of HCN and CN in the Orion Bar }

\author{M. Zannese\inst{1}
  \and
  J. R.\,Goicoechea\inst{1}
         \and
         S. Cuadrado\inst{1} 
        \and 
       J. Pety\inst{2,3}
       \and
       M. Gerin\inst{3}
       \and
       E. Chapillon\inst{2,4}
     }

\institute{Instituto de F\'{\i}sica Fundamental
     (CSIC). Calle Serrano 121-123, 28006, Madrid, Spain. \email{m.zannese@iff.csic.es}
\and
 Institut de Radioastronomie Millim\'etrique, 38406, 
 Saint Martin d’H$\grave{\rm e}$res, France.
\and
LUX, Observatoire de Paris, PSL Research University, CNRS, Sorbonne Universités, 75014 Paris, France.   
\and
Univ. Bordeaux, CNRS, Laboratoire d’Astrophysique de Bordeaux, UMR5804, 33600 Pessac, France.
}

   \date{Accepted 16/09/2026}

\abstract
{Rotationally excited HCN and CN lines, due to their high critical densities, are often used as tracers of gas densities. The Atacama Large (sub-) Millimeter Array (ALMA), with its high spectral and spatial resolution, provides unique access to the fundamental spatial morphology and kinematics of HCN and CN emission to study the substructures of the molecular cloud edges.}
{We present new sub-arcsecond imaging of HCN $4-3$ and CN $3-2$ lines to probe the density structure in the Orion Bar Photodissociation Region (PDR).}
{We study HCN $4-3$ and CN $3-2$ lines spatial distribution at very small scales (down to 0.5", $\simeq 200$~au) and compare them to other existing tracers inside the PDR, such as H$_2$ observed with JWST and C$_2$H observed with ALMA. We complemented these data with multiple $J$ lines observed with the IRAM 30m and \textit{Herschel} telescopes to study HCN and CN excitation near the dissociation front (DF). Finally, we compared observations with PDR and radiative transfer models to study nitrogen chemistry and to estimate gas density. }
{We find that HCN and CN emission are more spatially extended toward the more UV-shielded cloud than C$_2$H $4-3$ and H$_2$. However, they both peak very close to the DF, in disagreement with previous lower-angular-resolution observations of lower energy lines, which showed that HCN peaked deeper inside the PDR and traced colder gas. This morphology can be explained by the enhanced UV-driven chemistry. In very dense and irradiated environments, HCN is efficiently formed by the endoergic reaction H$_2$ + CN. Hence, its emission can also probe warmer gas near the DF. In addition, CN emission is observed in the atomic PDR -- where the PAH emission is strongest -- not accounted by stationary gas-phase chemistry due to the very low H$_2$ abundances.
 We interpret this as photoprocessing of CN-bearing PAHs. We find that the excitation of HCN and CN is subthermal $T_{\rm rot} \sim 20$~K (when the gas temperature is around $T_{\rm gas} \sim 200$~K), meaning that the gas cannot have densities much higher than the critical densities of the lines. We still derive high densities near the DF $n_{\rm H} \sim 10^6$~cm$^{-3}$ when comparing observations with \texttt{RADEX} and Meudon PDR Code models. Finally, we do not find evidence of small-scale ($\sim 100$~au, $\sim 10^{15}$~cm size), high-density clumps and argue that the observed density gradient could explain the variation of excitation between the different tracers. At the small spatial scales probed by ALMA, we resolve two velocity components in HCN emission: 10.5~km~s$^{-1}$ (the systemic velocity of the Bar) and 11.5~km~s$^{-1}$. The very structured emission of HCN observed in this redshifted emission component is more compatible with the signature of the propagation of a UV-induced shock compressing the gas than clumps. }
{This study highlights the bright HCN emission in warm layers of the PDR, challenging the traditional assumption that HCN primarily traces cold, dense gas.}

\maketitle

\section{Introduction}\label{sec:introduction}

Massive stars, by emitting an intense flux of far-ultraviolet (FUV) photons, influence the physical and chemical structure of their parent molecular clouds. They contribute to the dispersal of the cloud through gas photoevaporation and the addition of angular momentum, both of which have a direct impact on star formation. Hence, the edges of irradiated molecular clouds, called photodissociation regions \citep[PDRs; for a review, see e.g.,][]{Hollenbach_1999,Wolfire_2022}, are places where radiative feedback is dominant \citep[e.g.,][]{Inoguchi_2020}. Moreover, emission from PDRs, which reprocess a significant fraction of the radiation emitted by young massive stars, dominates the infrared (IR) spectra of the galaxies in which they are located. Studying these regions is therefore essential to better understand star formation and the evolution of interstellar matter. 

\begin{figure*}[th]
\centering   
\includegraphics[width=0.94\linewidth]{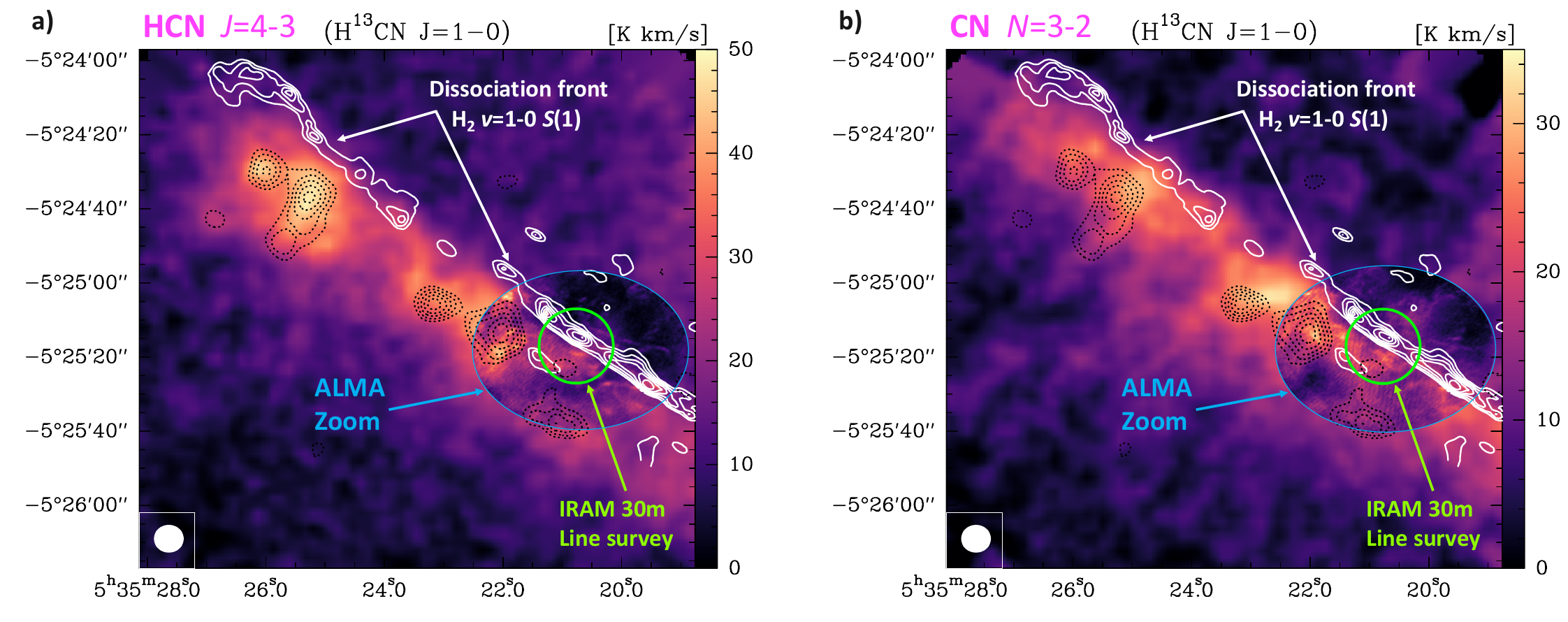}
\caption{Overview of the Orion Bar showing the FoV observed with ALMA and the
dissociation front (DF) position surveyed in depth with the IRAM\,30m telescope.
UV radiation from the Trapezium impinges from the upper right corner.
\mbox{\textbf{a)} HCN $J$\,=\,4--3} and  \mbox{\textbf{b)} CN $N$\,=\,3--2}  integrated line emission (color scale) mapped
with the IRAM\,30m telescope at $\sim$8$''$ angular resolution (white filled circle).
The ``ALMA zoom'' areas show the line emission observed by ALMA
at $\sim$0.5$''$ resolution (see an expanded view in Fig.~\ref{fig:MO_maps_ALMA}).  
The white  contours delineate the position of the H$_2$ DF traced by the H$_2$ $v$\,=\,1–0 $S$(1) line \citep[from 1.5 to \mbox{4.0\,$\times$\,10$^{-4}$\,erg\,s$^{-1}$\,cm$^{-2}$\,sr$^{-1}$} in steps of \mbox{0.5\,$\times$\,10$^{-4}$}; from][]{Walmsley00}.
The black dotted contours show large molecular clumps observed in
$\sim$5$''$ resolution interferometric observations of \mbox{H$^{13}$CN $J$\,=\,1--0}
 \citep[from][]{Lis_2003}.}
\label{fig:hcn_cn_e330}
\end{figure*}

Several observational studies (from IR to millimeter wavelengths) have suggested that the density structure in PDRs is inhomogeneous \citep[e.g.,][]{Tielens1993,vanderwerf_1993,Luhman_1998,Goicoechea_2016}. Indeed, when compared with PDR models, different species observed at the same position suggested different gas densities. For instance, isochoric PDR models require higher densities to reproduce the observed intensities of high-$J$ molecular lines (e.g., CO) than other species (e.g., C, C$^+$). It was thus proposed that the PDRs consist of high-density clumps ($n_{\rm H} \sim 10^6-10^7$~cm$^{-3}$) inside an interclump "low-density" medium  \citep[typically $n_{\rm H} \sim 10^3-10^5$~cm$^{-3}$,][]{Burton_1990,Parmar_1991,Meixner_1992,Hogerheijde_1995,Steiman_1997,Andree-Labsch_2017}. Using gas density tracers such as the HCN $1-0$ line, large clumps ($\sim 2000$~au, $\sim 3\times 10^{16}$~cm size) have been observed in PDRs \citep[e.g. in the Orion Bar,][]{Young_2000,Lis_2003}. However, until recently, the studies relied on low spatial resolution ($\lesssim 10$") observations, which were insufficient to resolve clumps at smaller scales  ($\sim 100$~au, $\sim 10^{15}$~cm size). The existence of these small clumps and their survival in irradiated environments have been theoretically studied in detail by \cite{Gorti_2002}. They reveal that these clumps undergo photoevaporation due to the ambient FUV field, which penetrates through their surface. Hence, many pressure-confined clumps (within an interclump medium) will quickly lose substantial fractions of their mass, thereby limiting the number of small clumps. In addition, more recent studies have shown that a density gradient in the PDR can naturally explain the observations without the need for a small-scale clumpiness hypothesis \citep[e.g.,][]{Marconi_1998,Habart_2005,Joblin_2018}. These observational studies were supported by photoevaporating PDR models \citep[e.g., \textit{Hydra},][]{Bron_2018}, which predict that the thermal pressure inside the PDR is almost constant (at most a factor 2 between the ionized and the atomic region), inducing a large density gradient (1 to 2 orders of magnitude between the atomic and the molecular region). The aim of this paper is therefore to investigate whether small high-density clumps or a density gradient exist using high-resolution observations.

To study the fundamental density structure of molecular clouds, HCN emission, widely used as a density tracer, can be a useful tool \citep[e.g.,][]{Santa_Maria_2023}. Indeed, HCN has a high dipole moment ($\upmu_e=2.99$~D), 30 times higher than that of CO. Hence, HCN lines (especially for higher-$J$ transitions) have high critical densities ($n_{\rm cr}(J=4-3) \sim 5 \times 10^6$~cm$^{-3}$ for collisions with H$_2$ at $T_{\rm k} = 100$~K, see Table \ref{tab:crit_dens}). Moreover, the CN/HCN abundance ratio can provide additional constraints, as its value is expected to increase near the stellar sources of the illuminating FUV radiation field \citep[e.g.,][]{Fuente_1993}, due to a more efficient photodissociation of HCN compared to CN \citep{van_zadelhoff_2003,Thi_2004}. Early extensive modeling of nitrogen chemistry in UV-irradiated environments showed that CN emission should trace the inner edges of the C/C$^+$ transition, whereas HCN will trace deeper, more shielded regions \citep{Boger_2005}.

The Orion Bar (see Fig. \ref{fig:hcn_cn_e330}), located in the Orion Molecular Cloud (OMC), the nearest massive star-forming region \citep[$d=414$ pc,][]{Menten_2007}, can be used  as a true interstellar laboratory for the study of density structure in PDRs. This nearly edge-on region is exposed to the intense FUV field from the Trapezium cluster, dominated by the O7-type star $\theta^1$ Ori C, with an effective temperature $T_{\rm eff} \simeq 40{,}000$~K. The intense FUV radiation field incident on the ionization front (IF) of the Bar is estimated to be $G_0 = 2-7 \times 10^4$ \citep[with $G_0 = 1$ corresponding to a flux integrated between 91.2 and 240 nm of $1.9 \times 10^{-3}$ erg cm$^{-2}$ s$^{-1}$,][]{Habing_1968} as derived from FUV-pumped oxygen IR-fluorescent lines \citep{Peeters_2024}. This intense FUV field shapes the edge of the cloud by ionizing and photodissociating the surrounding gas composing the Orion cloud. This produces the observed PDR, subdivided into several transitions layers between the \HII\ region, the neutral atomic layer and the molecular region. Previous maps of the low-lying of CN and HCN lines with the IRAM 30m telescope across the OMC probed the density and kinematic structure of the region at large scales \citep[e.g.,][]{Fuente_1996, Rodriguez-Franco_1998}. However, the relatively low spatial resolution of these observations ($10-30''$) did not allow the study of the very small-scale structure we investigate here.

In this paper, we present the analysis of new HCN and CN emission observed in the Orion Bar with the Atacama Large (sub)-Millimeter Array (ALMA) at very small scales (up to 0.5") as well as HCN and CN multiple $J$ lines obtained from the line surveys with IRAM 30m \citep{Cuadrado_2015,Cuadrado_2016,Cuadrado_2017,Cuadrado_2019} and \textit{Herschel}/HIFI \citep{Nagy_2017}. In Sect. \ref{sec:observations}, we present the ALMA and IRAM 30m observations and the data reduction. In Sect. \ref{sec:results}, we present the spatial morphology of HCN and CN emission at very small scales as well as the velocity components of their emission. We also study HCN and CN excitation thanks to the line survey. In Sect. \ref{sec:discussion}, we use HCN and CN emission to estimate the density throughout the Orion Bar using \texttt{RADEX} and the Meudon PDR Code. We also present a thorough analysis of HCN and CN chemistry under warm, irradiated conditions.

\section{Observations}\label{sec:observations}
\subsection{IRAM\,30m submm line maps}\label{sec:e330-obs}

To obtain a large-scale view of the Orion Bar, we mapped a $2.5' \times 2.5'$ region with the IRAM 30m telescope using the 330\,GHz E330 receiver (0.8\,mm atmospheric band) in combination with the FTS backend at 200\,kHz spectral resolution. \mbox{On-the-fly} (OTF) scans were obtained along and perpendicular to the Bar. The resulting spectra were gridded into a data cube by convolution with a Gaussian kernel, yielding a final resolution of $\sim$8$''$.
The total integration time was $\sim$6\,h under excellent winter conditions
($\lesssim$\,1\,mm of precipitable water vapor). The achieved rms noise is $\sim$1\,K per resolution channel. Figure~\ref{fig:hcn_cn_e330} shows the spatial distribution of the 
\mbox{HCN~$J$\,=\,$4-3$} (354.5\,GHz) and
\mbox{CN~$N$\,=\,$3-2$} (\mbox{$J$\,=\,7/2--5/2}, 340.2\,GHz) integrated line intensities.

\subsection{ALMA imaging of the Bar edge}\label{sec:alma-obs}

We carried out mosaics of a small field of the Orion Bar using forty-seven ALMA \mbox{12 m} antennas in \mbox{band 7} targeting the
\mbox{HCN~$J$\,=\,$4-3$} (354.5\,GHz) and
\mbox{CN~$N$\,=\,$3-2$, $J=7/2-5/2$} (340.25\,GHz, in the following termed as CN $3-2$) lines. 
These  observations belong to project 2015.1.01082.S (\mbox{P.I.: J. R. Goicoechea})
 and consisted of a 27-pointing mosaic centered at 
\mbox{$\alpha_{\rm J2000} = 5^{\rm h}35^{\rm m}20.50^{\rm s}$}; 
\mbox{$\delta_{\rm J2000} = -05\degree25'21.4''$}.
The usable field-of-view (FoV) is $\sim$40$''$$\times$40$''$. 
 
We observed the \mbox{HCN\,$J$\,=\,$4-3$} line and 
several \mbox{CN\,$N$\,=\,$3-2$}  hyperfine structure lines
using  correlators providing $\sim$500\,kHz resolution (0.4~km\,s$^{-1}$). The 
on-source observation time with the ALMA\,12\,m array 
was $\sim$1\,h. In order to recover the large-scale extended emission filtered out by the interferometer, we used deep and fully sampled single-dish maps, obtained with the total-power (TP) antennas at 19$''$ resolution,
as zero- and short-spacings. Data calibration procedures and image synthesis steps are described in \cite{Goicoechea_2016}. 
The synthesized beam is $\sim$0.5$''$.
This  is nearly a factor of $\sim$20 better than the previous
\mbox{HCN\,$J$\,=\,$1-0$} interferometric observations of the Orion Bar 
obtained with the Berkeley-Illinois-Maryland Association (BIMA) array \citep{Young_2000}
and a factor of $\sim$10 better than previous \mbox{H$^{13}$CN\,$J$\,=\,$1-0$} 
observations with the Plateau de Bure interferometer \citep{Lis_2003}.
The upper panels of Fig.~\ref{fig:MO_maps_ALMA} show the resulting
integrated line intensity images (\mbox{HCN~$J$\,=\,4--3} in the left panel
and the brightest \mbox{CN~$N$\,=\,3--2} (\mbox{$J$\,=\,7/2--5/2}) hfs lines at 340.247\,GHz in the right panel).
These images are rotated 54$^o$ clockwise to align the FUV illumination with the horizontal direction (from the right). 
The typical rms noise of the final cube is $\sim$\,0.4\,K at 354.5\,GHz ($\sim$\,0.3\,K at 340\,GHz)  per velocity channel and \mbox{0.5$''$-beam}. We compared these images to the C$_2$H $N=4-3$ line, which is presented and discussed in \cite{Goicoechea_2025}.
The lines observed with ALMA presented in this study are summarized in Table \ref{tab:crit_dens}. 

\begin{table}[!h]
  \begin{center}
    \caption{Spectroscopic information of the pure rotational lines studied in this work. 
   }
   
  \label{tab:crit_dens}
     \begin{tabular}{l   c c c@{\vrule height 10pt depth 5pt width 0pt}}   
      \hline \hline
      Species / Transition & Frequency  & $E_{\rm up}$/k   & $n_{\rm cr}^{\dag}$    \\ 
                           & $[$GHz$]$  &  $[$K$]$         &   $[$cm$^{-3}$$]$    \\         
      \hline
HCN\,\,\,\,$J=4-3$                & 354.505 &  42.5     & 5\,$\times$\,10$^6$ \tablefootmark{a} \\
 CN\,\,\,\,$N=3-2$                & 340.248 &  32.7     & 2\,$\times$\,10$^6$ \tablefootmark{b} \\
  C$_2$H\,\,\,\,$N=4-3$                & 349.338 &  41.9     & 3\,$\times$\,10$^6$\tablefootmark{c}\\\hline
      \end{tabular}
  \tablefoot{$^{\dag}$For collisions with H$_2$ at $T_{\rm k}$\,=\,100\,K, and assuming optically thin emission. Collisional data from \tablefoottext{a}{\cite{Hernandez_Vera_2017},}\tablefoottext{b}{\cite{Kalugina_2013},}\tablefoottext{c}{\cite{Dagdigian_2018}.}} 
  \end{center}
  
\end{table}

\subsection{IRAM\,30m and \textit{Herschel}/HIFI line surveys toward the DF}\label{sec:survey-obs}

We detected multiple rotational lines of HCN and CN using the  IRAM\,30m telescope (in Pico Veleta, Spain), toward the ``dissociation front'' (DF) position
at \mbox{$\mathrm{\alpha_{2000}=05^{h}\,35^{m}\,20.8^{s}\,}$}, 
\mbox{$\mathrm{\delta_{2000}=-\,05^{\circ}25'17.0''}$}
(green \mbox{circle} in \mbox{Fig.~\ref{fig:hcn_cn_e330}}, i.e., inside the field
observed with ALMA).
These observations are part of a complete line survey covering the frequency range \mbox{80\,$-$\,360\,GHz} \citep[i.e., 3, 2, 1, and 0.8\,mm atmospheric bands;][]{Cuadrado_2015,Cuadrado_2016,Cuadrado_2017,Cuadrado_2019}. 
 We used the EMIR receivers in combination with the Fast Fourier Transform Spectrometer (FTS) backends at 200\,kHz resolution (0.68\,km\,s$^{-1}$, 0.34\,km\,s$^{-1}$,
0.23\,km\,s$^{-1}$, and 0.17\,km\,s$^{-1}$ at $\sim$88\,GHz, $\sim$177\,GHz,
 $\sim$265\,GHz and $\sim$354\,GHz, respectively).  
We carried out these observations in the position switching mode taking a distant reference position at \mbox{($-$600$''$, 0$''$)}.  The half power beam width (HPBW) at $\sim$88\,GHz, $\sim$177\,GHz, $\sim$265\,GHz and $\sim$354\,GHz  is $\sim$28$''$, $\sim$14$''$, $\sim$9$''$ and $\sim$7$''$, respectively.  The latest observations (in the 2\,mm band) were conducted in \mbox{March 2020}. The data were first calibrated in the antenna temperature scale $T^{*}_{\rm A}$ and then  converted to the main beam temperature scale, $T_{\rm mb}$, using \mbox{$T_{\rm mb}$ = $T^{*}_{\rm A}/ \upeta_{\rm mb}$}, where $\upeta_{\rm mb}$ is the antenna efficiency. We reduced and analyzed the data using the GILDAS software, as described in \citet{Cuadrado_2015}. 
The typical rms noise is $\sim$ 3.3 mK (in antenna temperature units, $T^*_A$) per 0.66~km~s$^{-1}$ channel at 88.6 GHz, $\sim$ 18 mK ($\Delta v = 0.33$~km~s$^{-1}$) at 177.3 GHz, ~7.0 mK ($\Delta v = 0.22$~km~s$^{-1}$) at 265.9 GHz, and $\sim$ 34 mK ($\Delta v = 0.17$~km~s$^{-1}$) at 354.5 GHz. 
\mbox{Figure~\ref{fig:hcn_spectra30m}} (HCN) and
\mbox{Fig.~\ref{fig:cn_spectra30m}} (CN) show the detected rotational lines.

 We complemented our  dataset with higher frequency submm HCN and CN   lines 
 previously detected by \cite{Nagy_2017} with \mbox{\textit{Herschel Space Observatory}} 
 toward the so-called  \mbox{``CO$^+$ peak''} position \mbox{\citep{Stoerzer_1995}}.
This peak is located at only $\sim$4$''$ from the DF position (i.e., within the HPBW of these observations). These observations were carried out with the HIFI receiver \citep{deGraauw_2010} at a spectral-resolution of 1.1\,MHz (0.7\,km\,s$^{-1}$ at 500\,GHz). HIFI's HPBW 
 range from  $\sim$42$''$ to $\sim$20$''$
 in the \mbox{500\,-\,1000\,GHz}  window  \citep[][]{Roelfsema_2012}.
The list of complementary lines detected by Herschel includes 
\mbox{HCN~$J$\,=\,6--5} (531.7\,GHz) to \mbox{$J$\,=8--7} (708.9\,GHz), 
\mbox{CN~$N$\,=\,5--4} (566.9\,GHz) to \mbox{$N$\,=\,7--6} (793.5\,GHz). We used the  $W_{\rm mb}$ integrated line intensities (in K\,km\,s$^{-1}$)  shown in \mbox{Table A.1} of \mbox{\citet{Nagy_2017}}.

\section{Results}
\label{sec:results}
\subsection{Overview of the morphology of HCN and CN in the Orion Bar}

\begin{figure*}
    \centering
    \includegraphics[width=0.7\linewidth]{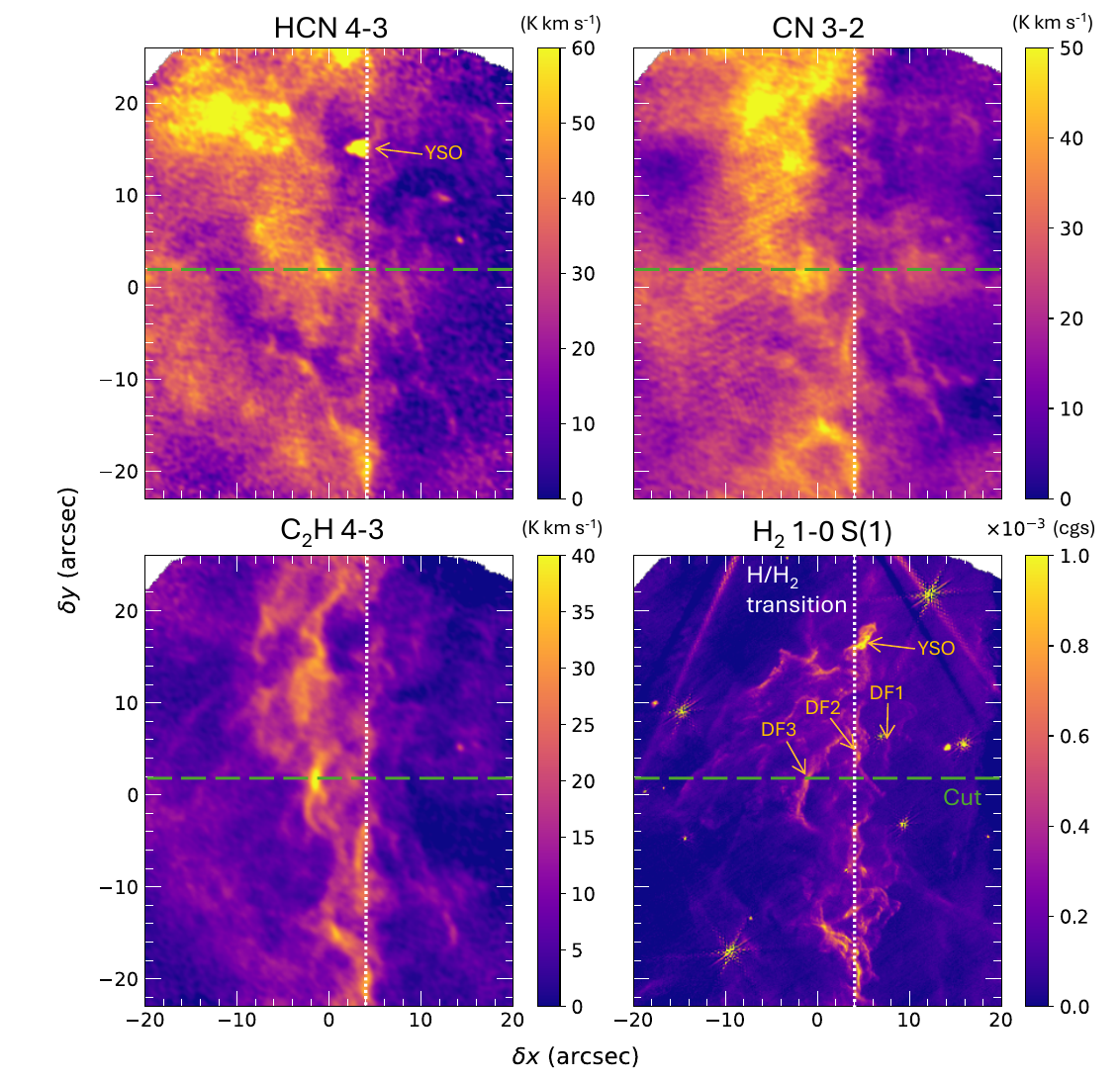}
\caption{Integrated line intensity maps of HCN $J=4-3$, CN $N=3-2$ $J=7/2-5/2$, C$_2$H $N=4-3$ $J=9/2-7/2$ obtained with ALMA and H$_2$ $1-0$ S(1) obtained with JWST as part of the GTO 1256 \citep{McCaughrean_2023}. The center of all images is at $\alpha_{\rm J2000} = 5^{\rm h}35^{\rm m}20.6^{\rm s}$; $\delta_{\rm J2000} = -05\degree25'20.0"$. All images have been rotated to have the front in the vertical direction and have the FUV-illuminating star in the horizontal direction (from the right). The vertical dashed white line represents the approximate position of the DF. The three DFs observed in the H$_2$ maps, as well as the Young Stellar Object (YSO) observed both in the H$_2$ map and the HCN map, are marked with yellow arrows.  The horizontal, dashed, green line indicates the spatial cut used for the rest of the study. The units “cgs” refer to erg~s$^{-1}$~cm$^{-2}$~sr$^{-1}$.}
    \label{fig:MO_maps_ALMA}
\end{figure*}
\begin{figure}
    \centering
    \includegraphics[width=\linewidth]{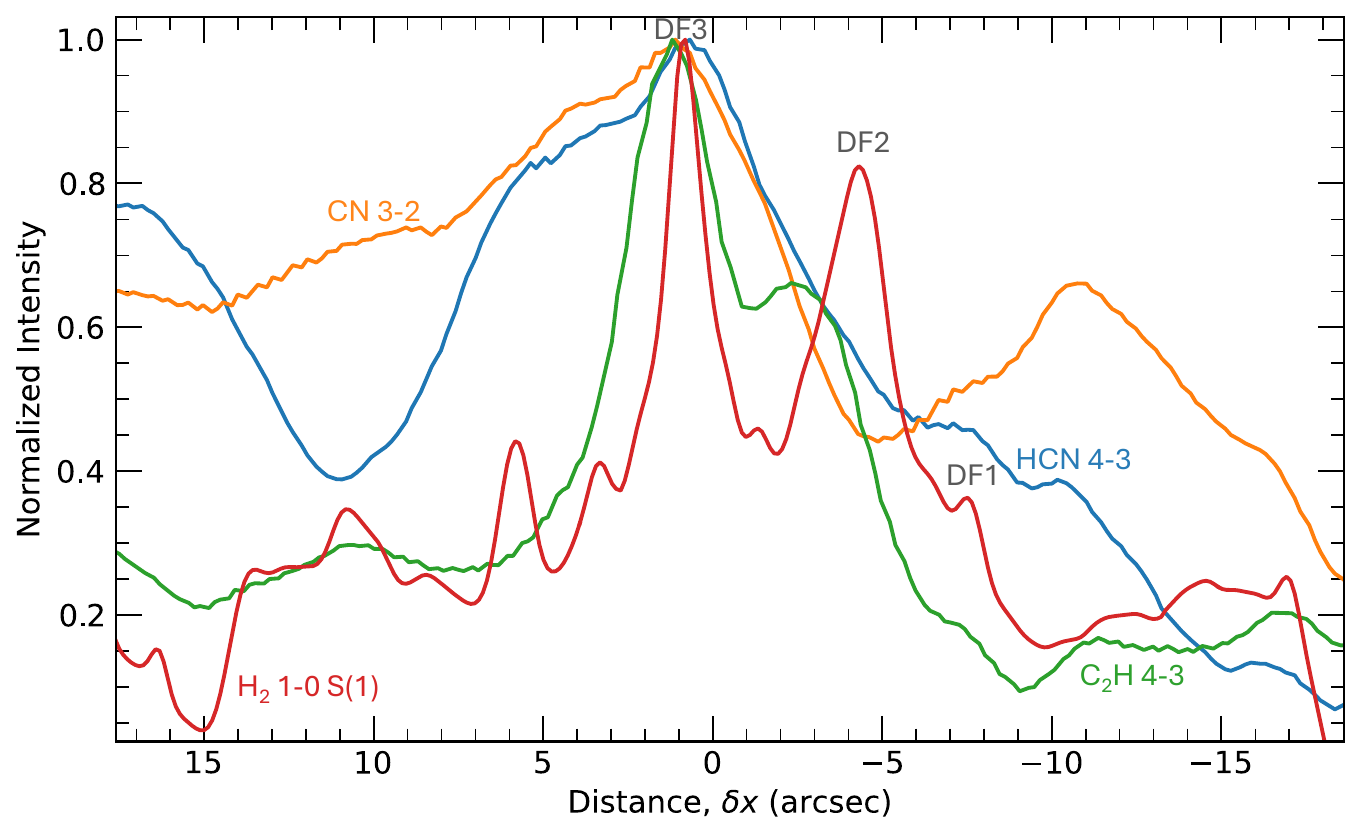}
    \caption{Normalized intensity profiles across the front (cut presented in Fig. \ref{fig:MO_maps_ALMA}) averaged over 4" perpendicular to the line cut. The illuminating star is located on the right.}
    \label{fig:spatial_cut}
\end{figure}
\begin{figure}
    \centering
    \includegraphics[width=0.75\linewidth]{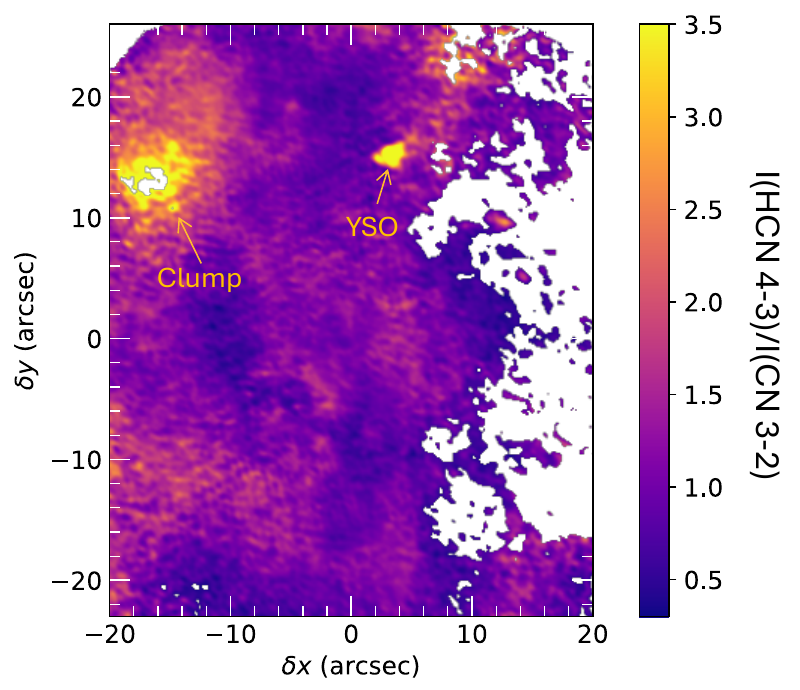}
    \caption{HCN $4-3$ / CN $3-2$ integrated line intensity ratio map. The ratio was calculated only where the lines were detected with at least 3$\sigma$.}
    \label{fig:map_ratio}
\end{figure}

\begin{figure*}
    \centering
    \includegraphics[width=\linewidth]{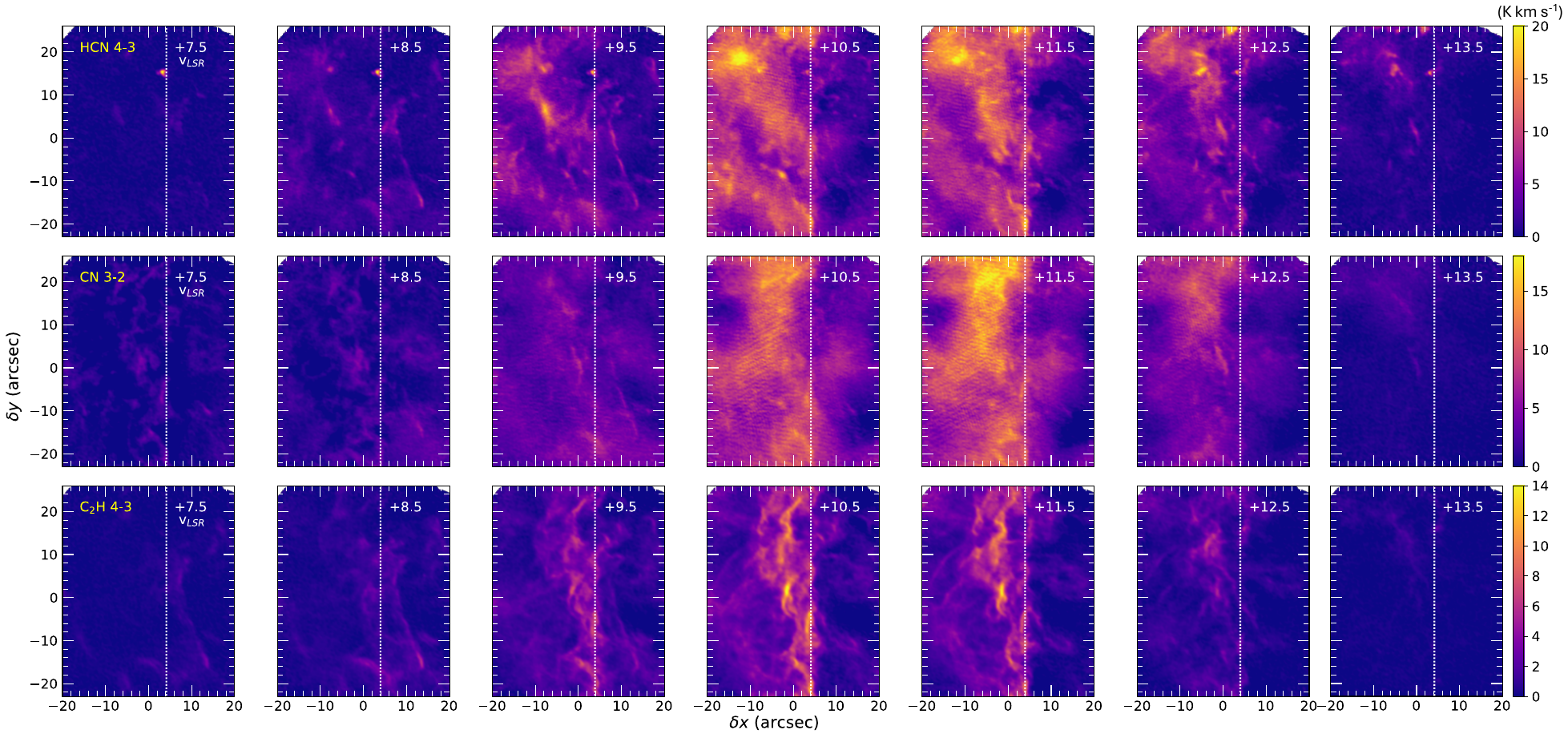}
    \caption{ALMA velocity channel maps of the Bar from $v_{ \rm LSR}=+7.5$ to $+13.5$~km~s$^{-1}$ in bins of 1~km~s$^{-1}$. The vertical dashed white line represents the approximate position of the DF.}
    \label{fig:velocity_maps}
\end{figure*}

The high spatial resolution obtained with ALMA observations allows us to study the morphology of HCN and CN emission at very small scales (0.5", $\simeq 200$~au). Figure \ref{fig:MO_maps_ALMA} displays the integrated intensity maps of the HCN $4-3$ line, CN $3-2$ line, C$_2$H $4-3$ line \citep[from][]{Goicoechea_2025} and H$_2$ $1-0$ S(1) line \citep[from \textit{James Webb Space Telescope} (JWST) data,][]{McCaughrean_2023}. This figure shows that, overall, HCN and CN have very similar morphology. The JWST H$_2$ image reveals three bright DFs \citep[DF1–DF3,][]{Habart_2024,Peeters_2024}, which we use as morphological references throughout this work. DF3 is the rim of the dense molecular gas, DF2 is probably a more translucent filament and DF1 is dominated by emission from the background emission from the background PDR at OMC-1 surface \citep[e.g.,][]{Goicoechea_2025}. The front edge of HCN and CN emission appears to be correlated with the H/H$_2$ DF, where C$_2$H and H$_2$ emissions peak. Nevertheless, their emission is more extended toward the more UV-shielded molecular cloud than the very narrow C$_2$H and H$_2$ emission. These results are highlighted in Fig. \ref{fig:spatial_cut}, displaying a spatial cut throughout the PDR. HCN and CN peak at the same position as C$_2$H and H$_2$ in DF3 (see Fig. \ref{fig:MO_maps_ALMA}). The main difference with these latter species is that they do not peak before DF3; their emission is rather low around DF2 and DF1. The peak of CN before DF1, in the atomic region ($\delta_x \sim 10.5"$), is not expected by standard, stationary  gas-phase chemistry due to the very low H$_2$ abundances. Its origin is further discussed in Sect. \ref{sect:CN-PAHs}. The second difference is that HCN and CN remain bright deeper into the PDR. Compared to the very structured H$_2$ and C$_2$H emission at the DFs, their emissions are more widespread and smoother, probably linked to a more complex chemistry, with various chemical pathways contributing to their formation. The spatial morphology we observe in HCN and CN emission with ALMA is in clear contradiction with the expected stratified structure, where CN peaks closer to the UV edge due to photodissociation of HCN \citep[e.g.,][]{Boger_2005,van_der_Wiel_2009}, while HCN peaks deeper in the UV-shielded gas due to its high critical density. 

\begin{figure*}
    \centering
    \includegraphics[width=\linewidth]{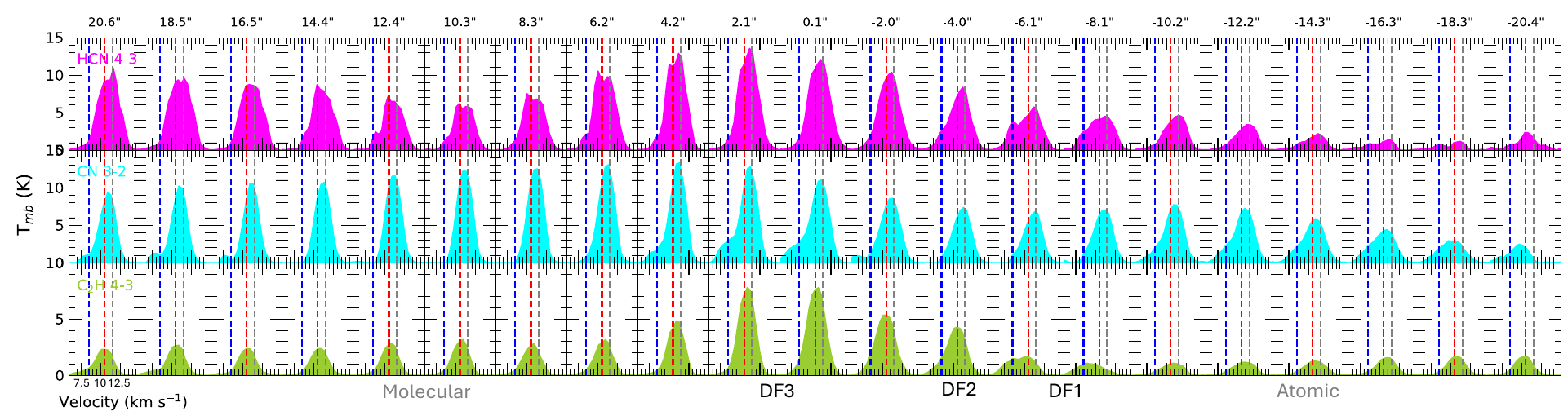}
    \caption{Spectra averaged over $2'' \times 2''$ boxes along the cut in the direction of the FUV-illuminating star (from the right). The dashed red lines mark the typical velocity centroid of the Bar, at $v_{\rm LSR} \simeq 10.5$~km~s$^{-1}$. The dashed blue lines mark the typical velocity centroid of the background OMC-1 emission at $v_{\rm LSR} \simeq 8.5$~km~s$^{-1}$. The dashed gray lines mark a velocity component at $v_{\rm LSR} \simeq 11.5$~km~s$^{-1}$.}
    \label{fig:profile_cut}
\end{figure*}

Still, there are a few differences between HCN and CN emissions which highlight physical conditions or chemical variations. For instance, a "river-like" structure is observed only in the HCN map ($\delta x = -10-0''$ and $\delta y= -10-0''$) where HCN is less bright, which could indicate a lower-density medium. In the top left part of the CN map ($\delta x = -16''$ and $\delta y= 10''$), we also observe a deficit of CN emission, which could be explained by the gas being more shielded from the UV field. Figure~\ref{fig:map_ratio} shows a map of the HCN $4-3$ / CN $3-2$ line intensity ratio. The ratio is relatively constant across the field of view, highlighting the strong correlation between HCN and CN emission. The differences noted above are more evident in this ratio. The HCN-to-CN ratio increases drastically in an extended zone at the top left ($\delta x = -16''$ and $\delta y= 15''$) and at the position of the Young Stellar Object (YSO, $\delta x = 3''$ and $\delta y= 15''$, Faraji et al. in prep), 
which could be associated with higher gas densities than the rest of the field of view. We further study the impact of the chemistry and physical conditions on HCN and CN emission in Sect. \ref{sec:discussion}.

\subsection{Velocity components of HCN and CN emission}

\label{sec:velocity}

Figure \ref{fig:velocity_maps} shows HCN $J=4-3$, CN $N=3-2$ and C$_2$H $N =4-3$ emission in different LSR velocity intervals. These plots dissect the line emission in 1~km~s$^{-1}$ channels, from $v_{\rm LSR} =+7.5$ to $+12.5$~km~s$^{-1}$. This figure shows that the majority of the emission originates from the component around $v_{\rm LSR} \sim 10-11$~km~s$^{-1}$ which corresponds to the systemic velocity of the Bar itself \citep[e.g.,][]{Cuadrado_2015} whereas the emission from the background OMC-1 (around $v_{\rm LSR} \sim 8-9$~km~s$^{-1}$), which is the brightest close to DF1, only contributes at a level almost an order of magnitude lower. This figure shows that HCN emission appears highly structured at redshifted velocities ($v_{\rm LSR} \sim 11.5-12.5$~km~s$^{-1}$), whereas CN is rather smooth at all velocities. Figure \ref{fig:profile_cut} displays velocity-resolved line profiles extracted across the cut presented in Fig. \ref{fig:MO_maps_ALMA}. These line profiles show that C$_2$H is mainly emitted at the Bar velocity around $v_{\rm LSR} \sim 10.5$~km~s$^{-1}$. On the contrary, CN seems to peak at slightly higher velocity around $v_{\rm LSR} \sim 11$~km~s$^{-1}$ and HCN has two clear components: one at $v_{\rm LSR} \sim 10.5$~km~s$^{-1}$ and one at $v_{\rm LSR} \sim 11.5$~km~s$^{-1}$. 
\begin{figure}
    \centering
    \includegraphics[width=\linewidth]{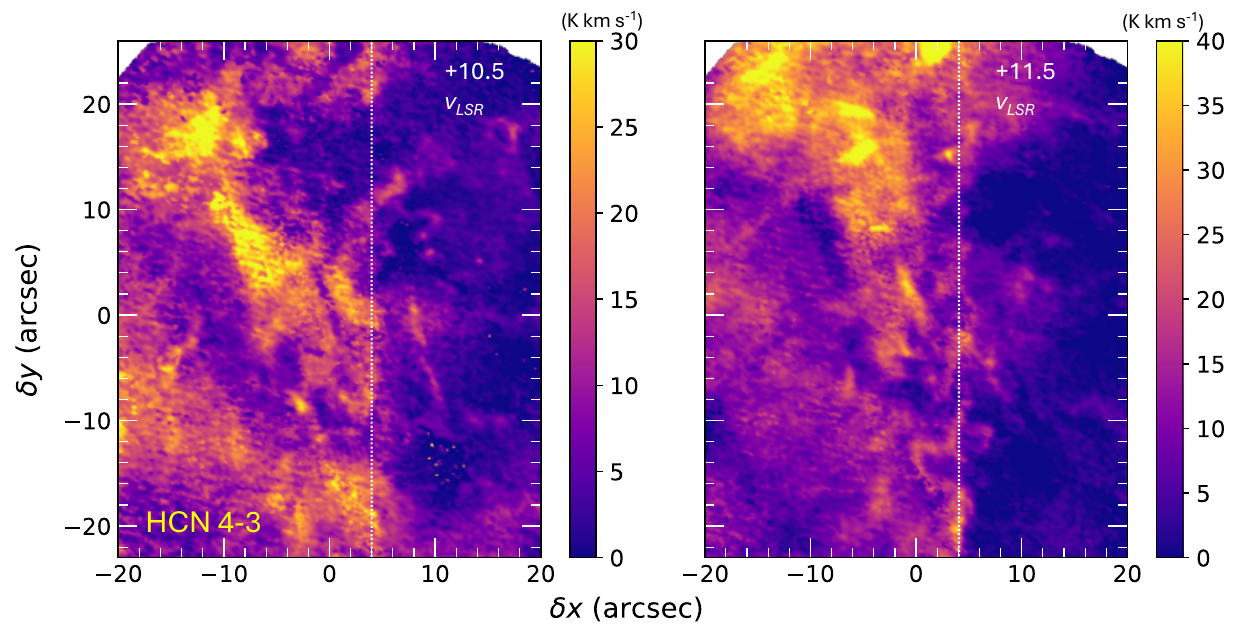}
    \caption{Integrated line intensities of the two velocity components of HCN $4-3$ emission in the Orion Bar at $v_{\rm LSR} = 10.5$~km~s$^{-1}$ and $v_{\rm LSR} = 11.5$~km~s$^{-1}$.}
    \label{fig:velocity_comp_gauss}
\end{figure}

To properly separate the observed components in HCN emission, we fit three Gaussians to the spectra at 8.5 (velocity of OMC-1), 10.5 (velocity of the Bar), and 11.5~km~s$^{-1}$ (observed redshifted component). Figure \ref{fig:velocity_comp_gauss} displays the fits of the 10.5 and 11.5~km~s$^{-1}$ components of HCN spectra. This figure shows that the two components trace different regions in the field of view. In the northeast part of the field of view, HCN emits at more redshifted velocities, whereas in the southwest  part, the emission is at the velocity of the Bar. This variation of velocity in the field of view has already been observed with the BIMA array in the HCN $1-0$ line \citep{Young_2000}. Interestingly, the northern east part is the brightest region in HCN integrated map (see Fig. \ref{fig:MO_maps_ALMA}) and 
a large bright clump in H$^{13}$CN $1-0$ is also observed at this position \citep{Lis_2003}. These results reveal that the dense clumps seem to be associated with redshifted velocities. This dense clump also seem bright at velocity between 11-12 km$^{-1}$ in CN $3-2$ emission (see Fig. \ref{fig:velocity_maps}). Moreover, \cite{Rodriguez-Franco_1998} observed CN $1-0$ emission with velocities between 11-12 km$^{-1}$ in the northeast part of the Bar and in more extended regions of Orion. Hence, these large structures observed with ALMA could be linked to a more extended complex kinematic observed in the cloud. In addition to these redshifted structures and thanks to the high spatial resolution of ALMA, we observe very structured filamentary HCN $4-3$ emission at small-scales at the dissociation front which originates from the 11.5~km~s$^{-1}$ component and not at the systemic velocity of the Bar. The emission of HCN at 10.5~km~s$^{-1}$ seems smoother near the DF. We further discuss the origin of these two components in Sect. \ref{sec:clump_filament}.

\subsection{Rotational excitation of HCN and CN}
\label{sect:diag_rot}
\begin{figure}
    \centering
    \includegraphics[width=0.75\linewidth]{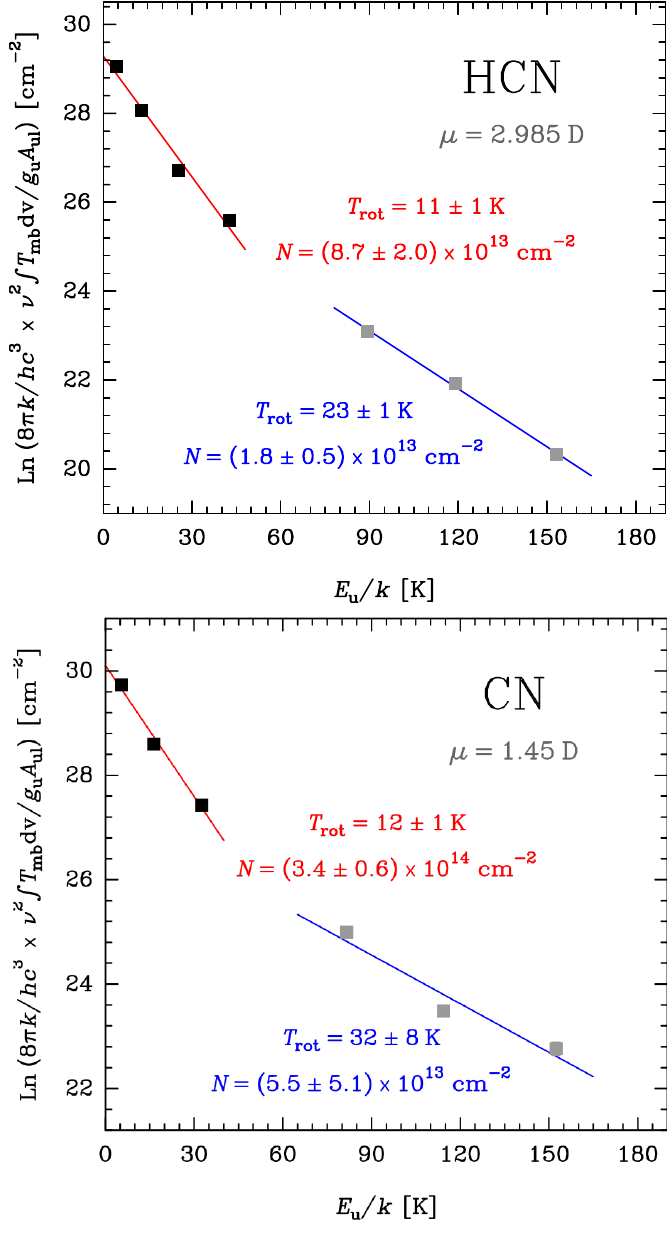}
\caption{HCN and CN rotational diagrams determined from observations conducted with the IRAM\,30m and \textit{Herschel}/HIFI \citep[taken from][]{Nagy_2017} telescopes toward the Orion Bar DF. (Top panel) HCN ($J$\,=\,6--5 to 8--7 lines observed with HIFI; gray points). (Bottom panel) CN ($N$\,=\,5--4 to 7--6 lines observed with HIFI). We corrected the line intensities with the frequency-dependent beam-coupling factors ($f_{\rm b}$) calculated in  Appendix \ref{sec:correction-obs}. Each diagram shows the fitted column density ($N$) and rotational temperature ($T_{\rm rot}$), along with their uncertainties.} 
    \label{fig:diag_rot}
\end{figure}

The HCN and CN rotational lines detected in the IRAM 30m and \textit{Herschel}/HIFI line surveys allow us to study the excitation of HCN and CN close to the DF. The parameters of HCN and CN lines are summarized in Table \ref{Table_HCN} and \ref{Table_CN}. Figure \ref{fig:diag_rot} displays the rotational diagram of HCN and CN.  The diagrams both show two rotational-temperature components:  a lower-excitation component ($T_{\rm rot~low} = 11$~K for HCN and $T_{\rm rot~low} = 12$~K for CN) and a higher-excitation component ($T_{\rm rot~high} = 23$~K for HCN and $T_{\rm rot~high} = 32$~K for CN). Both these components are subthermal ($T_{\rm rot} \ll T_{\rm gas}$). Indeed, at the DF, H$_2$ observations indicate a temperature around 600~K \citep[][Zannese et al., submitted]{van_de_putte_2024}. In addition, CO $3-2$ observations obtained with ALMA, which, due to its optical thickness and low critical density, is a very good tracer of gas temperature, point to a temperature $\gtrsim 200$~K near the DF \citep{Goicoechea_2016}. This HCN and CN subthermal excitation implies that the gas density cannot be a lot higher than the critical densities (see Table~\ref{tab:crit_dens}). Furthermore, the lines are moderately optically thick ($\tau \lesssim 4$\footnote{Here, the opacity is defined at the center of the rotational line, neglecting all fine and hyperfine components, which are not spectrally resolved in the HCN $4-3$ emission detected with ALMA.}) and radiative trapping becomes important as lines become optically thick, leading to lower effective critical densities $n_{\rm cr,eff} \sim \frac{n_{\rm cr}}{\tau}$ \citep[e.g.,][]{Evans_1999,Shirley_2015}. Hence, we expect the density cannot be a lot higher than 10$^6$~cm$^{-3}$.

As the lines are in the effectively thin regime (meaning the line intensities still scale with column density, see Appendix \ref{opacity} and discussion below) and assuming they follow the beam coupling factor (calculated in Appendix \ref{sec:correction-obs}), we can, as a first approximation, derive the column density of these molecules $N$ from the excitation diagram using the Boltzmann distribution equation:

\begin{equation}
\ln\left(\frac{N_{\rm up}}{g_{\rm up}}\right)= \ln\left(\frac{N}{Q(T_{\rm rot})}\right)-\frac{E_{\rm up}}{k_{\rm B} T_{\rm rot}},
\label{eq:excit_diagram}
\end{equation}
where $N_{\rm up}$ is the column density of the upper level of the studied transition, $g_{\rm up}$ the upper-level degeneracy,   $T_{\rm rot}$ the excitation temperature, $Q(T_{\rm rot})$ the partition function and $k_{\rm B}$ the Boltzmann constant. As HCN and CN distributions suggest two temperature components, we can estimate the total column density as $N \simeq N_{\rm rot~low} + N_{\rm rot~high}$.

In the DF, we recover comparable column densities of HCN and CN. $N$(HCN) $ \sim 10^{14}$~cm$^{-2}$ and $N$(CN) $ \sim 4\times10^{14}$~cm$^{-2}$. These column densities, however, are uncertain for various reasons. First of all, they deeply depend on the estimation of the beam coupling factor. Indeed, these factors are calculated assuming that every line of HCN and CN follows the morphology of HCN $4-3$ and CN $3-2$ emission, respectively. However, the lower excitation lines could have more extended emission. This would lead to an overestimation of the beam-corrected intensity of the lower excited lines and, consequently, the column density. To estimate the uncertainty on the derived column densities, we also consider the values derived without correcting for beam dilution, which would give a lower limit. Hence, the lower limits are $N_{\rm low}$(HCN) $ = (3.4 \pm 0.2) \times 10^{13}$~cm$^{-2}$ and $N_{\rm low}$(CN) $ = (2.5 \pm 0.3) \times 10^{14}$~cm$^{-2}$. The second source of uncertainty is the opacity of the lines. As discussed before, at these column densities, HCN and CN lines become optically thick (see Fig. \ref{fig:prop_I_Ncol} which displays the intensity of the studied lines as a function of the column density of HCN and CN). However, Eq. (\ref{eq:excit_diagram}) is only valid for optically thin lines. Yet, these lines are still moderately optically thick ($\tau~\lesssim~4$), so owing to a subthermal excitation, they fall in the effectively thin regime. Hence, the column density derived from the excitation diagram is expected to remain close to the real column density but could be slightly underestimated by less than a factor of 2. 

\section{Discussion}
\label{sec:discussion}
\subsection{Estimation of the gas density}
\subsubsection{Modeling with RADEX}
\label{sect:radex}

\begin{figure}
    \centering
    \includegraphics[width=0.95\linewidth]{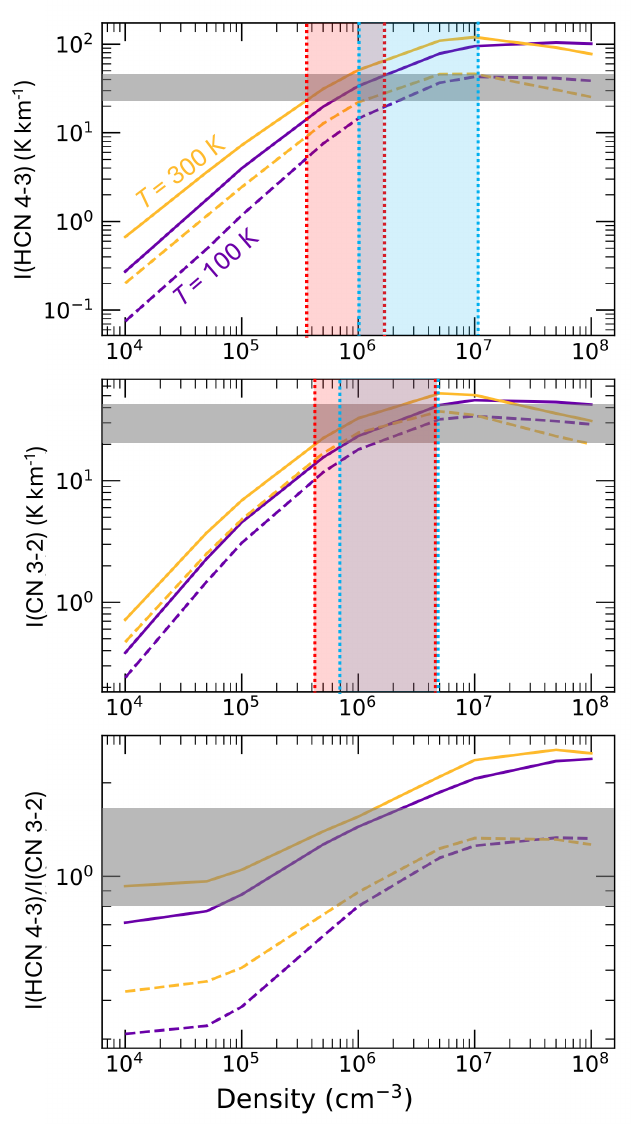}
    \caption{Prediction of the line intensity of HCN $4-3$ (top panel), CN $3-2$ (middle panel) and the ratio of these lines (bottom panel) by \texttt{RADEX} models for different gas temperatures as a function of gas density. The solid lines are for the observed column density considering beam dilution factors ($N$(HCN) $ = 1.1 \times 10^{14}$~cm$^{-2}$ and $N$(CN) $ = 4.0 \times 10^{14}$~cm$^{-2}$) and the dashed lines are for the not corrected column densities ($N$(HCN) $ = 3.4 \times 10^{13}$~cm$^{-2}$ and $N$(CN) $ = 2.5 \times 10^{14}$~cm$^{-2}$). The grayed area corresponds to the range of intensities and intensity ratios observed near the DF3 in ALMA data. The red (resp. blue) area corresponds to the density range constrained by the models, considering the corrected (resp. uncorrected) column densities.}
    \label{fig:radex_density}
\end{figure}

We used the \texttt{RADEX} code \citep{van_der_tak_2013}, which computes the non-LTE excitation and the radiative transfer in the escape probability approximation in a single zone, to predict HCN and CN rotational line intensities as a function of temperature and density. The hyperfine structure of HCN and the fine structure of CN is not resolved in our calculation. Using latest data resolving fine and hyperfine structure \citep[only up to $T_{\rm gas} = 100$~K,][]{Kalugina_2015}, we estimate that the observed component $I$(CN $J=7/2-5/2$) (in this paper named $I$(CN $3-2$) for the sake of simplicity) is about half of the total rotational line intensity. We implemented the latest inelastic collisional data available (HCN-$e^-$ \citep{Faure_2007}, HCN-H$_2$ \citep{Hernandez_Vera_2017}, CN-$e^-$ \citep{Harrisson_2013}, CN-H$_2$ \citep{Kalugina_2013}). We produced a grid of models with gas densities ranging from $n_{\rm H} = 10^4 - 10^8$~cm$^{-3}$. We modeled two gas temperatures $T=100$~K and $T=300$~K, based on the CO $3-2$ observations, which point to a temperature around $T\sim 200$~K \citep{Goicoechea_2016}. As discussed in Sect. \ref{sect:diag_rot}, the column densities derived from the single-dish observations are uncertain because of the optical thickness of lines and the uncertainty in the estimation of the beam coupling factor (due to the different HPBW of the observations). Hence we chose two extreme values: one from the corrected intensities ($N$(HCN) $ = 1.1 \times 10^{14}$~cm$^{-2}$ and $N$(CN) $ = 4.0 \times 10^{14}$~cm$^{-2}$) and one from uncorrected intensities ($N_{\rm low}$(HCN) $ = 3.4 \times 10^{13}$~cm$^{-2}$ and $N_{\rm low}$(CN) $ = 2.5 \times 10^{14}$~cm$^{-2}$). We adopt a line width $\Delta v = 2.5$~km~s$^{-1}$ based on the width of C$_2$H $4-3$ emission. Indeed, we expect HCN $4-3$ and CN $3-2$ to have similar line widths, as they originate from similar regions, but their hyperfine and fine-structure blending prevents us from precisely measuring the Doppler broadening.  As \texttt{RADEX} takes as an input the density of the collisional partners, we assume the density of H$_2$ $n_{\rm H_2} = n_{\rm H}/2$, where $n_{\rm H} = n(\text{H}) +2n(\text{H}_2)$, and we assume that the H$_2$ ortho-to-para ratio is thermalized to the gas temperature. In the rotationally excited HCN and CN lines, electrons have a less important impact than H$_2$ molecules in the collisions because the critical fractional abundance of electrons needed to dominate the collisional excitation, $x_{{\rm cr},\,e^-} = n_{\rm cr}(e^-) / n_{\rm cr}$(H$_2$) 
\citep[e.g.,][]{Goldsmith17,Santa_Maria_2023}, is higher than the abundance of electrons. Still, we consider them with  $n(e^{-}) \simeq n(\text{C}^+) =  10^{-4} n_{\rm H}$.  As \texttt{RADEX} also considers radiative pumping, we chose the intensity of the IR background radiation field using the spectrum observed with the JWST with the PDRs4All program\footnote{DOI: 10.17909/pg4c-1737} \citep{ERS_2022,Peeters_2024} and a modified blackbody at $T = 100$~K with a spectral index $\beta =2$.

Fig. \ref{fig:radex_density} shows the \texttt{RADEX} intensity predictions of HCN, CN and their ratio as a function of gas density. Comparing these predictions with the total integrated intensities of HCN and CN, ignoring velocity components, near the DFs (more precisely just before and after DF3), allows us to constrain the local density. Using the predictions calculated with the beam-filling corrected column densities, HCN observations point to densities around $n_{\rm H} = 4 \times 10^5 - 2 \times 10^6$~cm$^{-3}$ and CN observations point to slightly higher but consistent values $n_{\rm H} = 4 \times 10^5 - 5 \times 10^6$~cm$^{-3}$. Using the uncorrected column densities yields higher density values, similar for both HCN and CN, in the range $n_{\rm H} = 7 \times 10^5~-~10^7$~cm$^{-3}$. The density could be smaller by at most a factor of 2, considering that we observe two distinct velocity components in HCN $4-3$ emission, which have approximatively the same strength (see Fig. \ref{fig:profile_cut} and \ref{fig:velocity_comp_gauss}). Once the column densities are fixed, the HCN~$4-3$/CN~$3-2$ intensity ratio does not provide strong additional constraints on the gas density (values ranging from $10^4$ to $10^8$~cm$^{-3}$ are compatible with the observed ratio). However, the previously determined density range is consistent with the observed ratio.

\begin{figure}
    \centering
    \includegraphics[width=0.95\linewidth]{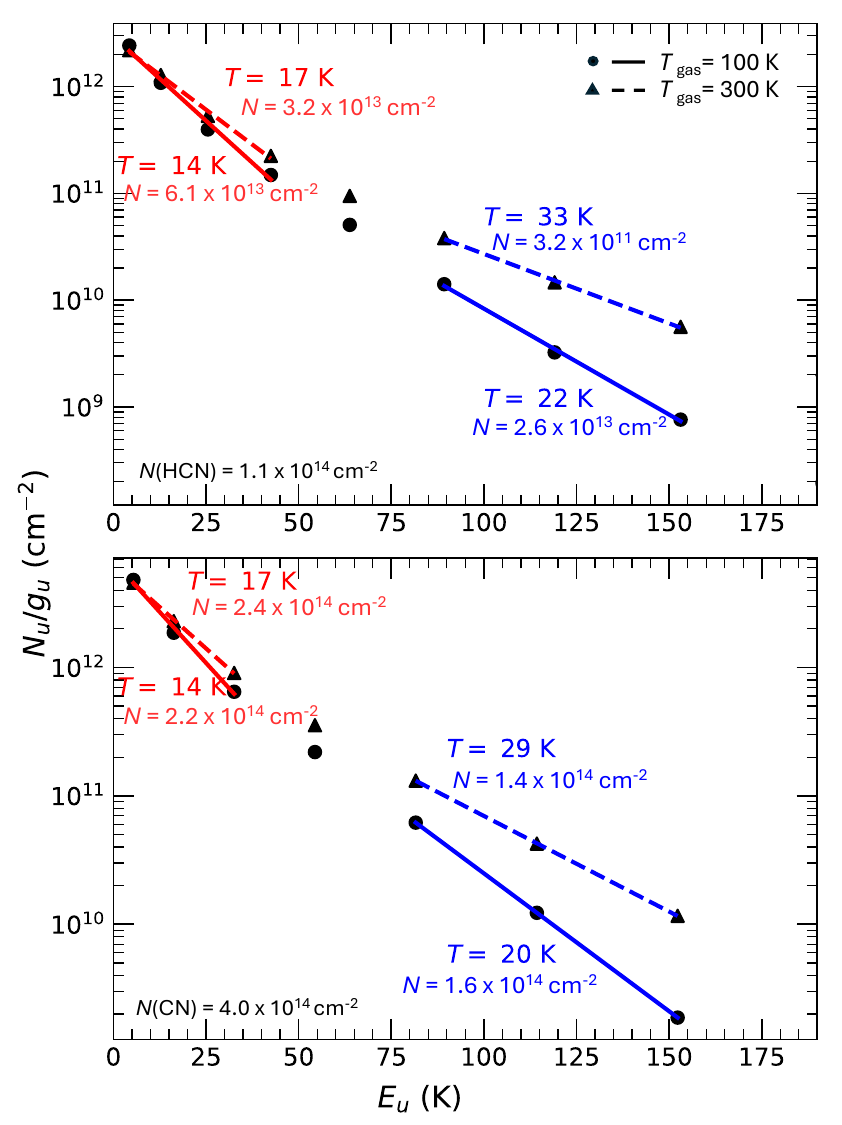}
    \caption{Rotational diagram of HCN (top panel) and CN (bottom panel) obtained with \texttt{RADEX} models at $n_{\rm H}=10^6$~cm$^{-3}$ using the observed column densities $N$(HCN) $ = 1.1 \times 10^{14}$~cm$^{-2}$ and $N$(CN) $ = 4.0 \times 10^{14}$~cm$^{-2}$. The points and solid lines correspond to a model at $T_{\rm gas} = 100$~K and the triangles and dashed lines correspond to a model at $T_{\rm gas} = 300$~K. The fits are made on the same levels as the observations (see Fig. \ref{fig:diag_rot}).}
    \label{fig:diag_rot_radex}
\end{figure}

Fig. \ref{fig:diag_rot_radex}
displays the excitation diagram of a model at $n_{\rm H} = 10^6$~cm$^{-3}$, which is in the range of what was previously derived, for two gas temperatures $T_{\rm gas} = 100$~K and $T_{\rm gas}=300$~K. We recover distinct rotational-temperature components in the excitation diagram. The first one is around 15~K for both molecules and both gas temperatures, and the second ranges from $T_{\rm rot}=20$~K to $T_{\rm rot}=30$~K, depending on the gas temperature. The lower-excitation components are slightly higher than observed, suggesting a possible overestimation of the $1-0$ beam-corrected intensity due to beam dilution factors. The higher-excitation HCN observed component, $T_{\rm rot~high} = 23$~K, is consistent with the model at $T_{\rm gas} = 100$~K, while the higher-excitation CN observed component $T_{\rm rot~high} = 32$~K is consistent with the model at $T_{\rm gas} = 300$~K. This is consistent with the fact that CN mainly traces the warm UV-irradiated layers, whereas HCN can also trace the colder, denser layers.

\subsubsection{Modeling with the Meudon PDR Code}
\label{sec:model_meudon}

With \texttt{RADEX}, we estimated the gas density from the intensity of HCN and CN lines from a single-zone model. However, we needed an estimation of the column density which, as discussed before, is uncertain and could vary across the field of view. Hence, we chose to produce a grid of Meudon PDR Code models \citep[version 7\footnote{Adapted from the public version of 2024: \url{https://pdr.obspm.fr/pdr_download.html}},][]{Le_Petit_2006} to independently constrain the gas density. The code simulates the thermal and chemical structure of the gas in a self-consistent manner, assuming a 1D geometry and a stationary, plane-parallel UV-irradiated gas-dust layer. Here, we used the prescription of \cite{Draine_1978} for the incident interstellar radiation field with a scaling of $G_0 = 10^4$. The code includes the progressive attenuation of the UV field due to grain and gas extinction \citep{Goicoechea_2007}. In this work, we assumed the extinction curve HD38087 of \cite{Fitzpatrick_1990} and $R_V =5.5$.
Compared with previous PDR models, here we used an updated version of the Meudon PDR code that includes the photoionization of vibrationally excited H$_2$ and state-dependent reactions between H$_2$($v$) and H$^+$ \citep{Goicoechea_2025b}. These processes increase the extent of the atomic layer of the PDR. HCN and CN emission near the DFs originate from a very thin layer where the density is almost constant (see Sect. \ref{sec:form_des}, Fig. \ref{fig:model_p1e8}). In addition, ALMA beam (0.5") mostly samples a parcel of gas with approximately constant density. Hence, to constrain the local density in a similar manner as the single-zone models presented in Sect. \ref{sect:radex}, isochoric models are sufficient to reproduce the observed intensity locally. We computed a grid of isochoric models ranging from $n_{\rm H} = 10^3-10^7$~cm$^{-3}$ with $G_0 = 10^4$, up to $A_V = 5$ to only trace the hotter layer of the PDR.
\begin{figure}
    \centering
    \includegraphics[width=0.9\linewidth]{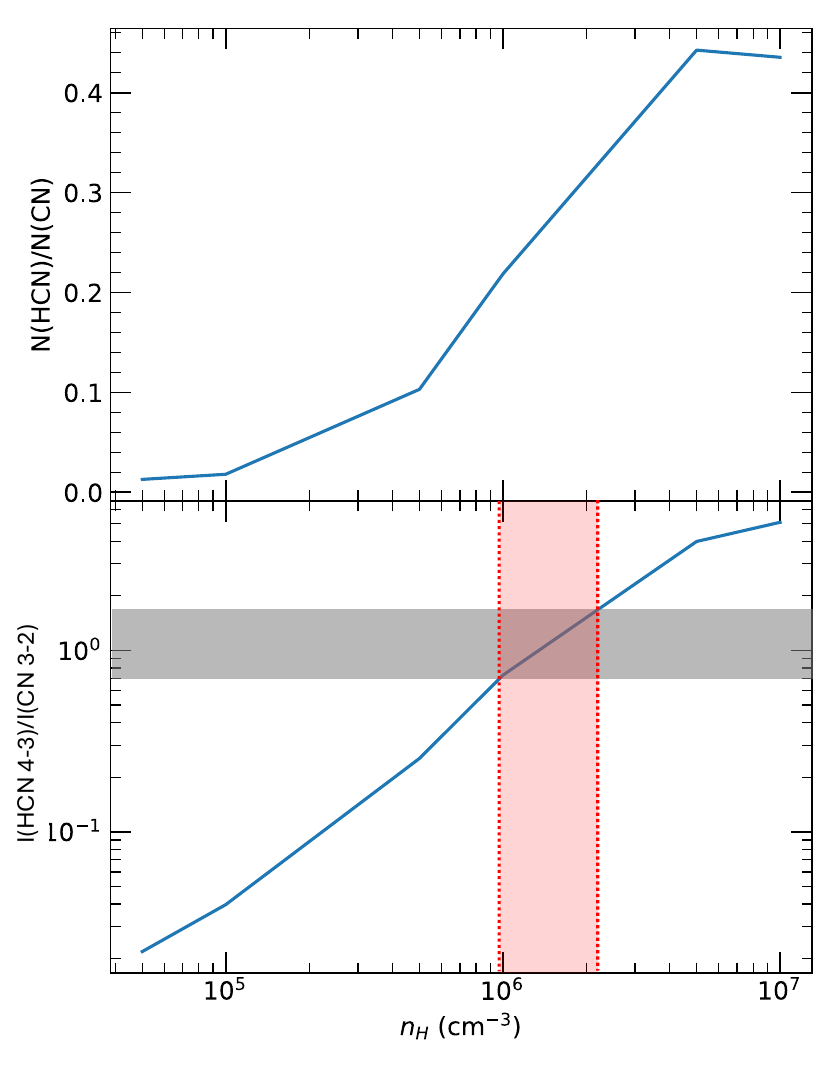}
    \caption{Meudon PDR code predictions for the HCN/CN column density ratio (top) and line intensity ratio (bottom)  as a function of density for models at $G_0 = 10^4$ and $A_V = 5$. The grayed area corresponds to the range of HCN/CN intensity ratios observed near the DF3 in ALMA data. The red area corresponds to the constrained gas density range.}
    \label{fig:ratio_N_I_meudon}
\end{figure}
Comparing observed absolute line intensities to the Meudon PDR Code outputs is not straightforward, as the geometry (inclination) of the PDR will impact the observed column density. However, Fig. \ref{fig:MO_maps_ALMA} revealed that HCN and CN emission trace similar regions, thus their line intensity ratios should not depend on the geometry of the PDR if their emission is effectively thin. Figure \ref{fig:ratio_N_I_meudon} displays the evolution of the column densities and intensities ratio as a function of density. The top panel shows that the ratio of column densities depends strongly on gas density. Nevertheless, comparing the observed value of intensity ratios to the predicted ones displayed in the bottom panel of Fig. \ref{fig:ratio_N_I_meudon} yields a very similar value of density as with the single-zone models around $n_{\rm H} = (1-2) \times 10^6$~cm$^{-3}$. This result confirms that in the emitting regions of HCN and CN, near the DF, the gas density is high, with $n_{\rm H} = 10^6$~cm$^{-3}$. As the gas temperature is also high at this position $T_{\rm gas} \sim 200$~K, the thermal pressure is thus around $P_{\rm th}/k_B \sim 2 \times 10^8$~K~cm$^{-3}$. This value is in agreement with previous estimation made with \textit{Herschel} observations \citep[from CO, OH, CH$^+$ lines,][]{Joblin_2018} but also values derived from CH$^+$ emission detected with the JWST \citep{Zannese_2025}. It is also in agreement with a pressure gradient near the DF as observations of H$_2$ and HD, 
both emission peaking before the H/H$_2$ transition, point to lower thermal pressure around $P_{\rm th}/k_B = 5\times 10^7$~K~cm$^{-3}$ \citep[Meshaka et al. in prep,][]{Zannese_2026}.
\subsection{High density clumps or compressed filaments?}
\label{sec:clump_filament}
Several studies have used lower-angular resolution HCN observations to infer the existence of high-density "big" clumps  ($\sim 5''$, $\sim 2000$~au) inside the Orion Bar \citep[e.g.,][]{Young_2000,Lis_2003}. In the HCN $4-3$ map (see Fig. \ref{fig:MO_maps_ALMA}), we find the presence of a large clump in the northern east part of the observations ($\delta_x \sim -15''$" and $\delta_y \sim 15$"), which was previously observed in H$^{13}$CN $1-0$ emission \citep{Lis_2003} and HCN $1-0$ emission \citep{Young_2000}. The ratio of HCN and CN intensities, which we have proved is a good tracer of density, shows that at this position the density must be significantly higher than the rest of the field of view.  

Previous low-angular resolution studies revealed spatially extended fine-structure line emission of C and C$^+$ near \HII\ regions, indicating that the FUV penetrates deeper into the cloud than predicted by homogeneous models \citep{Stutzki_1988,Meixner_1992,Steiman_1997}, suggesting that the gas is clumpy, probably at small scales. Our high spatial resolution ALMA observations also allow us to study the density structure of the PDR at very small scales and thus study the presence of small clumps, traced by HCN emission, near the DF. However, our study shows that in the field of view, except from the large clump and the YSO, the density is not expected to vary much and is still high $n_{\rm H} \sim 10^6$~cm$^{-3}$, so a lot higher than what was previously expected for the "interclump" medium $n_{\rm interclump} \sim 10^4-10^5$~cm$^{-3}$ \citep{Young_2000}. The overall structure of the Orion Bar after the DF, traced by HCN emission, looks rather homogeneous. This result is consistent with the analysis of \cite{Gorti_2002}, which shows that small clumps within an interclump medium are quickly photoevaporated and lose substantial fractions of their mass in very short time scales.

Certainly, the velocity maps of the HCN $4-3$ emission do look structured (see Fig. \ref{fig:velocity_maps}). However, we observe two velocity components in HCN $4-3$ emission (see Sect. \ref{sec:velocity}) and Fig. \ref{fig:velocity_comp_gauss} shows that this highly structured emission 
originates from the second component, at a higher velocity than quiescent gas in the Bar, and is more filamentary than globular. On the contrary, the emission at the Bar's systemic velocity looks smoother. The question is what is the origin of these structures. These filamentary structures appear perpendicular to the DFs and have width around 1", hence very close to what is observed for the H$_2$ filaments (see Fig. \ref{fig:MO_maps_ALMA} and Zannese et al. submitted). Thus, this seems compatible with compressed substructures produced by the propagation of a UV-driven shock wave due to the expansion of the HII region and/or the photoevaporation of the PDR. Previous works have studied how the flow of gas (advection, photoevaporation...) through PDRs can impact their structure \citep[e.g.,][]{Bertoldi_1996,Bron_2018,Maillard_2021}. The ionization of the surface layer of the PDR creates a shock that propagates inside the PDR ahead of the DF. \cite{Bertoldi_1996} estimates a shock velocity of $v_s = 6.5$~km~s$^{-1}$ in Orion A (see their Eq. (37)). \cite{Bron_2018} find a lower value $v_s = 3$~km~s$^{-1}$  when modeling a photoevaporating PDR with the \textit{Hydra} PDR code for representative parameters for the Orion Bar, $n_{\rm H} = 10^5$~cm$^{-3}$ and $G_0 = 2 \times 10^4$. In their grid of models, they find shock velocity between 0.5 and 7~km~s$^{-1}$. Here, we find two different gaussian components with a shift in velocity between the redshifted component and the velocity of the Bar of the order of 1~km~s$^{-1}$. If the velocity shift is associated with the signature of the propagation of a shock compressing the gas, then, as the Orion Bar is observed almost edge-on, the absolute shock velocity would be $v_s = \frac{\Delta v}{cos(i)}$ where $i$ is the angle between the observer and the normal to the PDR. If we assume around $i = 80 \degree$, then the absolute shock velocity is $v_s \simeq 6$~km~s$^{-1}$, which would be consistent with the previous estimations.

These results, the structure of the PDR traced by HCN~$4~-~3$ emission not appearing clumpy, the recovery of high density everywhere in the field of view, and the shift between the two velocity components in HCN $4-3$ emission being compatible with a UV-driven shock in a photoevaporating PDR, are consistent with a non-clumpy medium with filamentary compressed substructures. In addition, as mentioned previously, we find lower density in the atomic region $n_{\rm H} = 5\times 10^4 - 10^5$~cm$^{-3}$ from molecular emission \citep[H$_2$, HD...,][]{Zannese_2026}. This density gradient could account for the density contrasts required to excite the different tracers, without the need for small-scale clumps previously invoked when only low to moderate-angular resolution observations existed.

\subsection{Modeling nitrogen chemistry}
\subsubsection{Main formation and destruction pathways}
\label{sec:form_des}
\begin{figure}
    \centering
    \includegraphics[width=\linewidth]{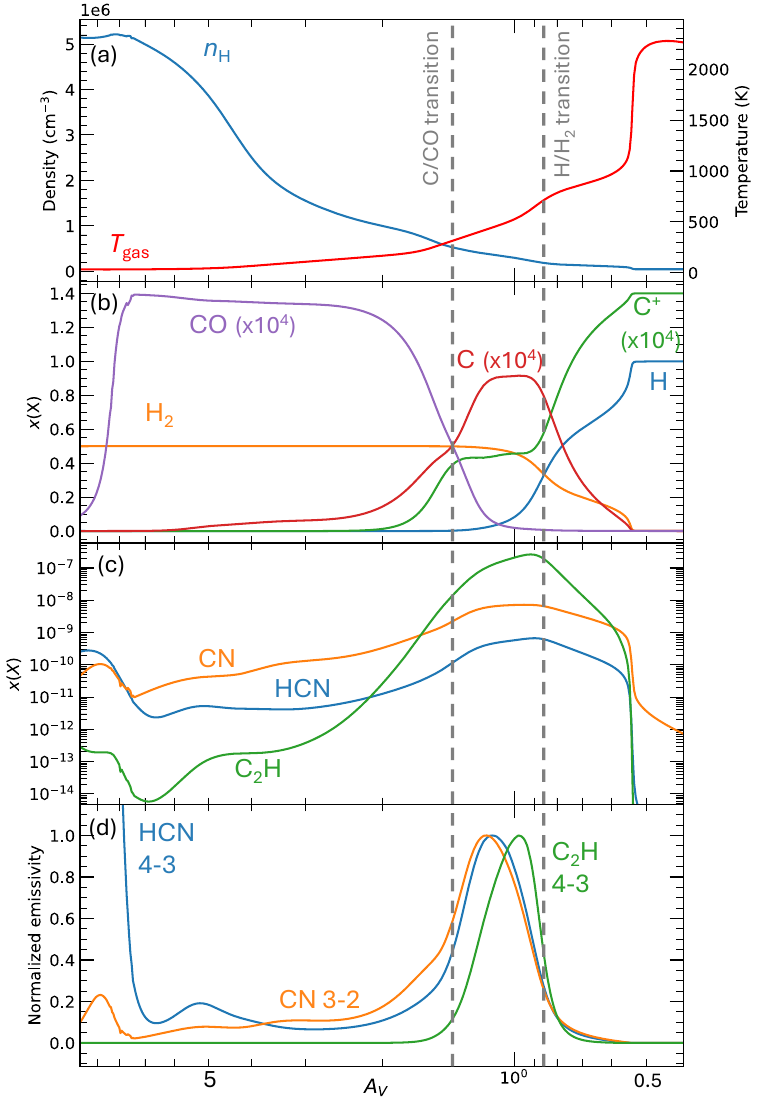}
    \caption{Meudon PDR model at $P_{\rm th}/k_B = 10^8$~K cm$^{-3}$ and $G_0 = 2 \times 10^4$. (a) Density and temperature profile across the PDR. (b) Abundances with respect to proton density of the main species. The abundances of carbon species are multiplied by 10$^4$ for representation purposes. (c) Normalized (on the first peak) abundances of the studied species: HCN, CN, and C$_2$H. (d) Normalized (on the first peak) emissivity of the studied lines: HCN $J=4-3$, CN $N=3-2$ and C$_2$H $N=4-3$. The dashed gray lines mark the position of the H/H$_2$ transition and C/CO transition.}
    \label{fig:model_p1e8}
\end{figure}

\begin{figure}
    \centering
    \includegraphics[width=0.75\linewidth]{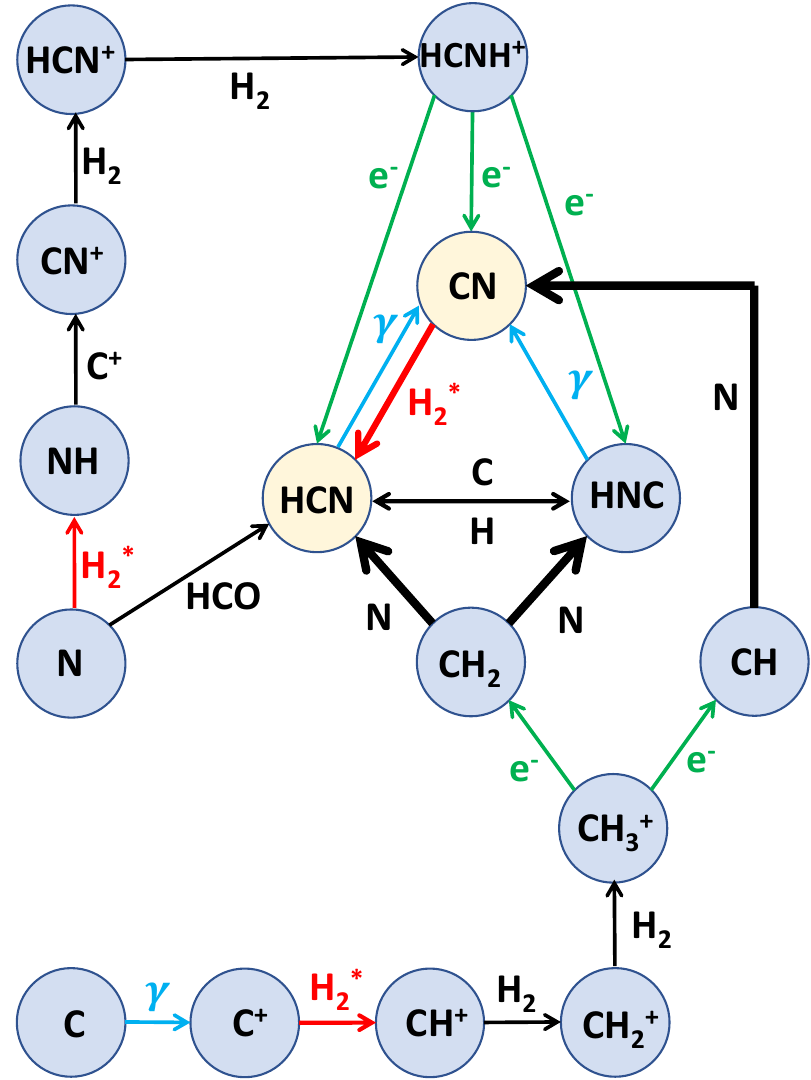}
    \caption{Dominant chemical reactions in FUV-illuminated gas. Red arrows indicate endoergic reactions, which proceed rapidly at high $T$ or in regions with a significant fraction of excited H$_2^*$.}
    \label{fig:chem_network}
\end{figure}

To analyze the origins of HCN and CN emission across all layers of the PDR, we use a fiducial model representative of the Orion Bar from the Meudon PDR code. We adopt an isobaric PDR with a thermal pressure of $P_{\rm th}/k_B = 10^8$~K cm$^{-3}$ and an intensity of the incident UV field $G_0 = 2 \times 10^4$, which are the parameters which can best reproduce other tracers in the Orion Bar \citep[e.g.,][]{Joblin_2018,Goicoechea_2025}. Figure \ref{fig:model_p1e8} displays the main physical parameters (gas temperature and density) and the abundances of the main species as a function of the depth (in units of visual extinction) inside the PDR. The bottom panels show the normalized abundances of HCN, CN, and C$_2$H, and the emissivities of the transitions studied in this paper: HCN $4-3$, CN $3-2$, and C$_2$H $4-3$. This figure shows that at high thermal pressure, HCN and CN peak very close to the H/H$_2$ transition and before the C/CO transition, slightly deeper than C$_2$H. This is in agreement with the observations as presented before (see Fig. \ref{fig:spatial_cut}). In this model, HCN and CN emissions peak toward the DF but become fainter in the more shielded regions similarly to C$_2$H, which is inconsistent with the observations. This difference could be explained by different hypotheses. First, the PDR is not a 1D slab but a 3D complex structure. If the medium is sufficiently porous, FUV radiation can penetrate through the intermediate Av layers and activate some UV-induced chemistry \citep[e.g.,][]{Gaches_2026}. In addition, it might not be truly isobaric, and a pressure gradient could be expected. In the fiducial model, owing to the steep density gradient, the gas temperature drops quickly after the H/H$_2$ transition, whereas in the observations, the temperature at this position remains high. This could explain why HCN and CN remain bright after the DF, as they are still efficiently formed by warm gas-phase chemistry. The network of dominant chemical reactions of this warm chemistry expected at $A_V<4$ is presented in Fig. \ref{fig:chem_network}. 

\begin{figure}
    \centering
\includegraphics[width=\linewidth]{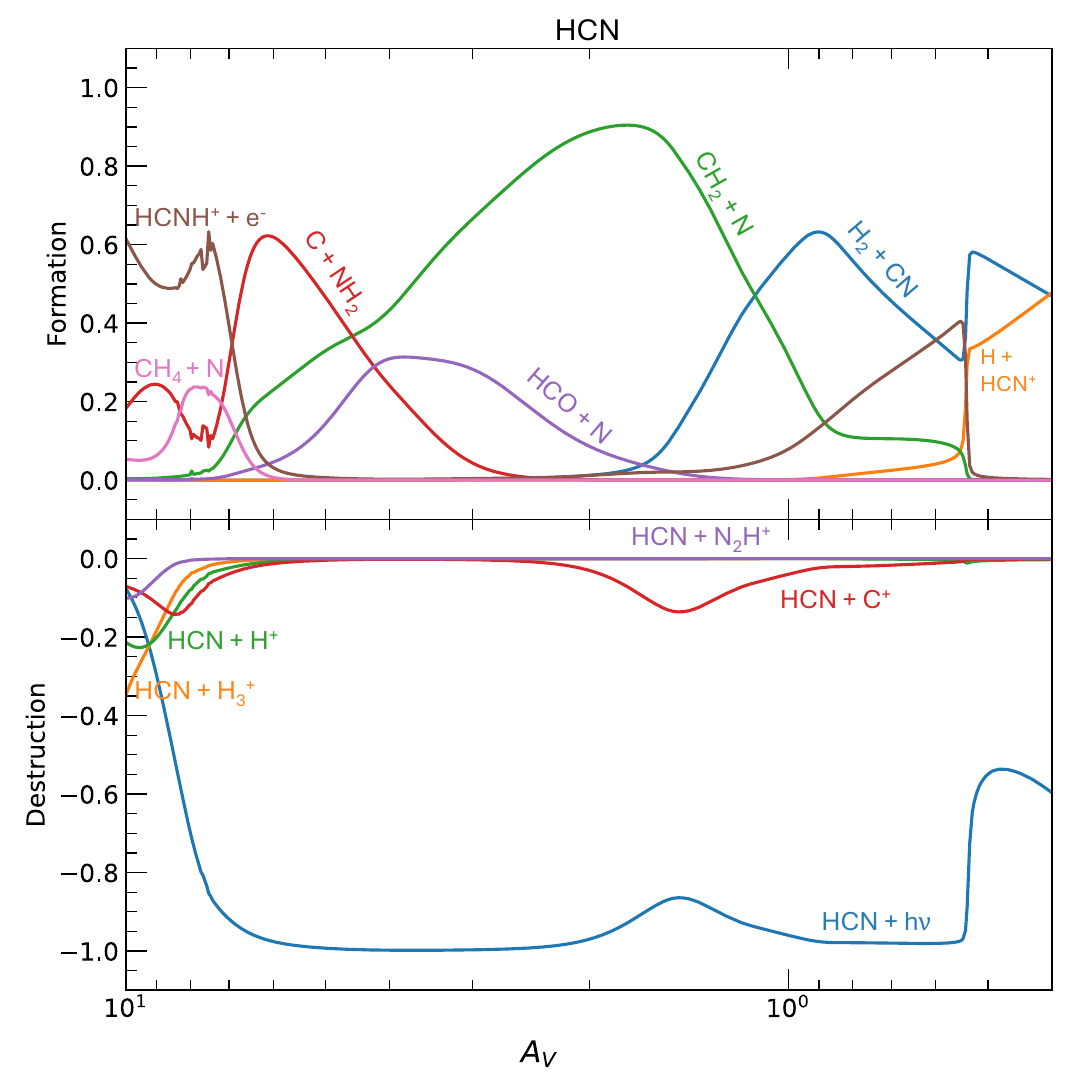}
\includegraphics[width=\linewidth]{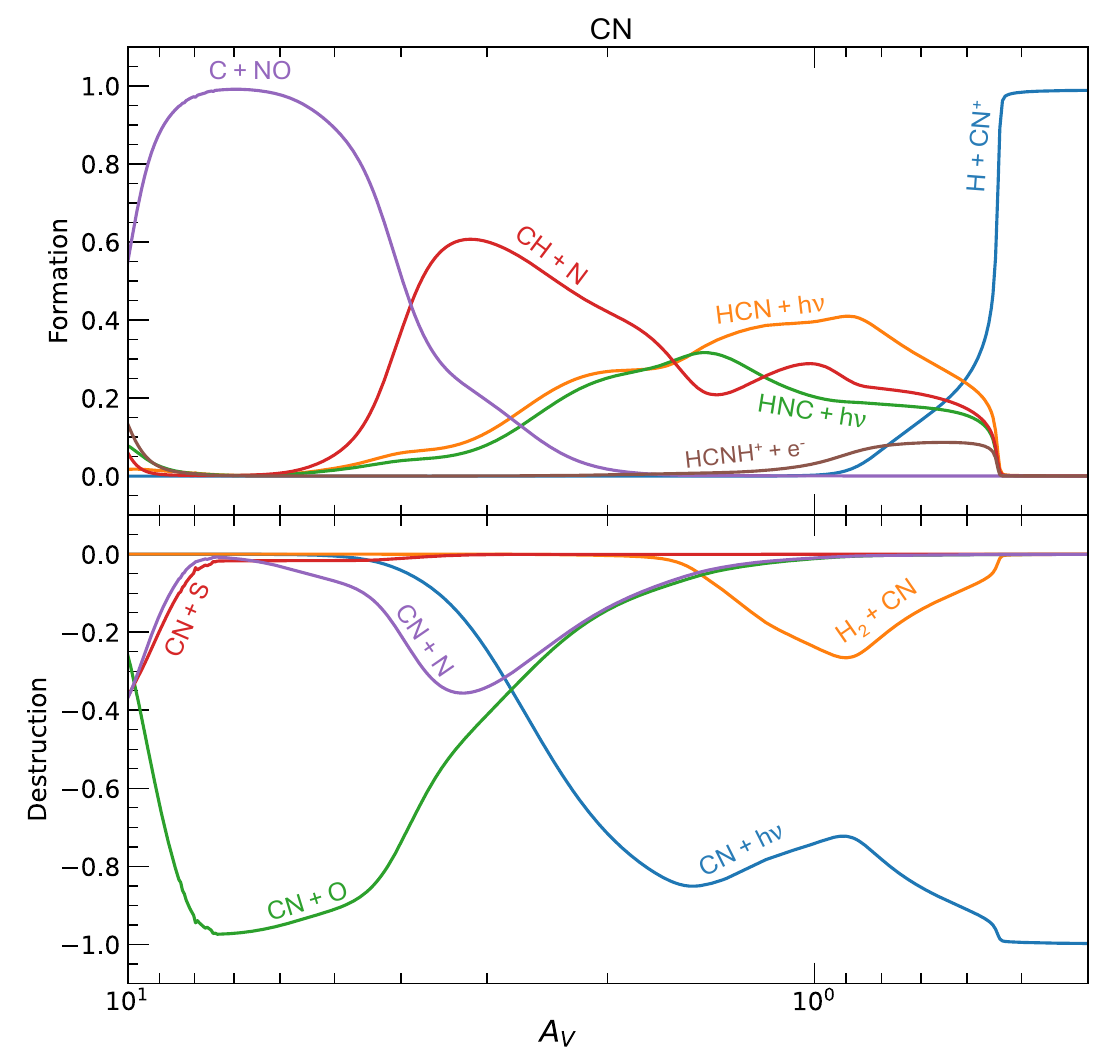}
    \caption{Main formation and destruction pathways contribution for (top) HCN and (bottom) CN as a function of depth into the PDR for a model at $P_{\rm th}/k_B = 10^8$~K cm$^{-3}$ and $G_0 = 2 \times 10^4$.}
    \label{fig:model_chemistry}
\end{figure}

Figure \ref{fig:model_chemistry} displays the main formation and destruction pathways of HCN and CN as a function of $A_V$ for our fiducial model. The chemistry of HCN and CN is quite complex with several pathways to form the molecules at both low and high $A_V$ \citep[e.g.,][]{Fuente_1993,Boger_2005}. Close to the edge, the chemistry is dominated by warm-neutral chemistry initiated by UV irradiation and reactions with warm or FUV-pumped H$_2^*$. Until $A_V \sim 2$, HCN is mainly formed by:
\begin{equation}
    \ce{H_2 + CN \rightarrow HCN + H}
\end{equation}
which is slightly endoergic $\Delta E = 820$~K \citep{Baulch_1994,Baulch_2005}. Even deeper into the PDR (until $A_V \sim 5)$, HCN formation is still closely indirectly linked to the warm H$_2$ induced chemistry as it is formed with the small hydrocarbon CH$_2$ \citep{Herbst_2000}:
\begin{equation}
    \ce{CH_2 + N \rightarrow HCN + H}
\end{equation}
Indeed, CH$_2$ \citep[detected in the Orion Bar,][]{Jacob_2021} forms in the gas phase from the electronic recombination of CH$_3^+$, detected in the Bar \citep{Zannese_2025}, which itself is the product of a chain reaction with H$_2$ starting with C$^+$ (see Fig. \ref{fig:chem_network}). CN chemistry is also enhanced by UV irradiation near the edge of the PDR as it is formed by the photodestruction of HCN and HNC \citep[e.g.,][]{Heays_2017}, and deeper into the PDR through reactions involving the hydrocarbon CH \citep[e.g.,][]{Brownsword1996,Daranlot_2013,Loison_2014}:
\begin{equation}
    \ce{CH + N \rightarrow CN + H}.
\end{equation}
CH was also detected in the Bar by \cite{Nagy_2017}, and similarly to CH$_2$, it is produced by the electronic recombination of CH$_3^+$ (see Fig. \ref{fig:chem_network}). The primary destruction pathway for both HCN and CN is photodissociation. Hence, at the edge of the PDR, the UV-driven reactions govern the chemistry of HCN and CN.

\subsubsection{Evolution of the chemistry with physical conditions}
\begin{figure}
    \centering
    \includegraphics[width=\linewidth]{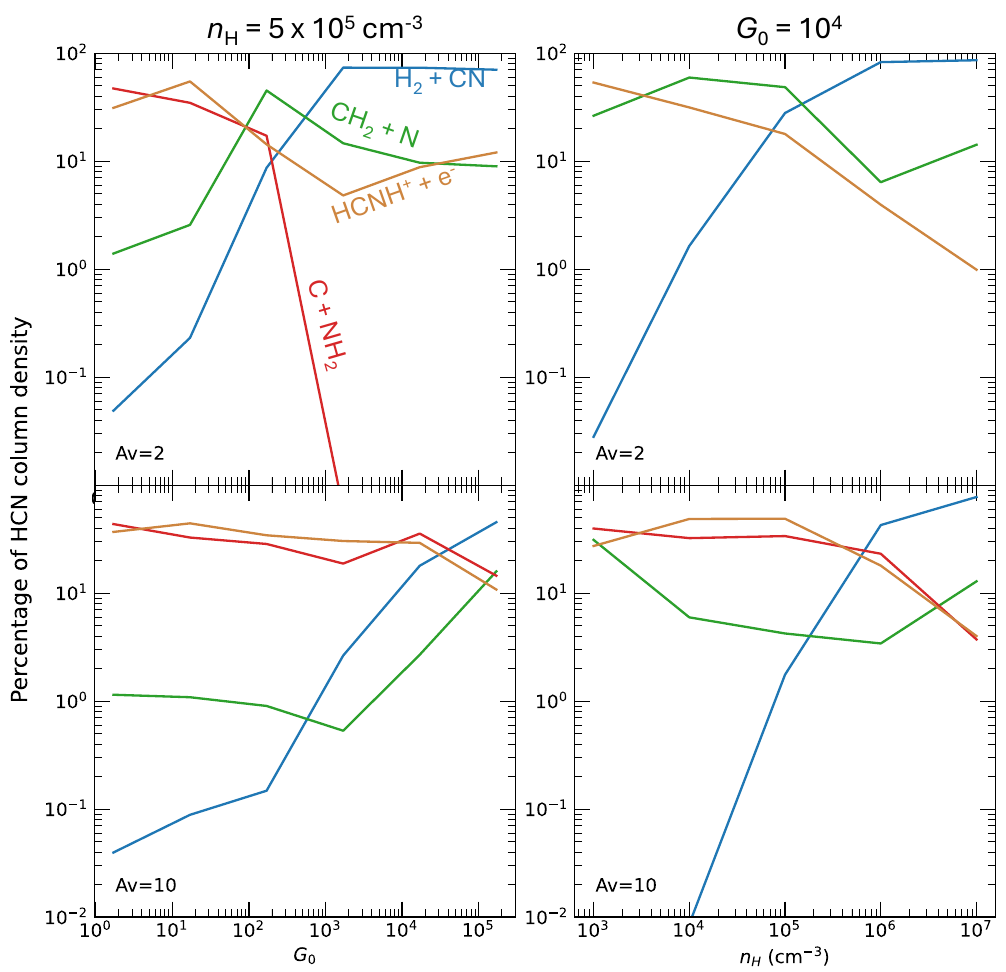}
    
    \includegraphics[width=\linewidth]{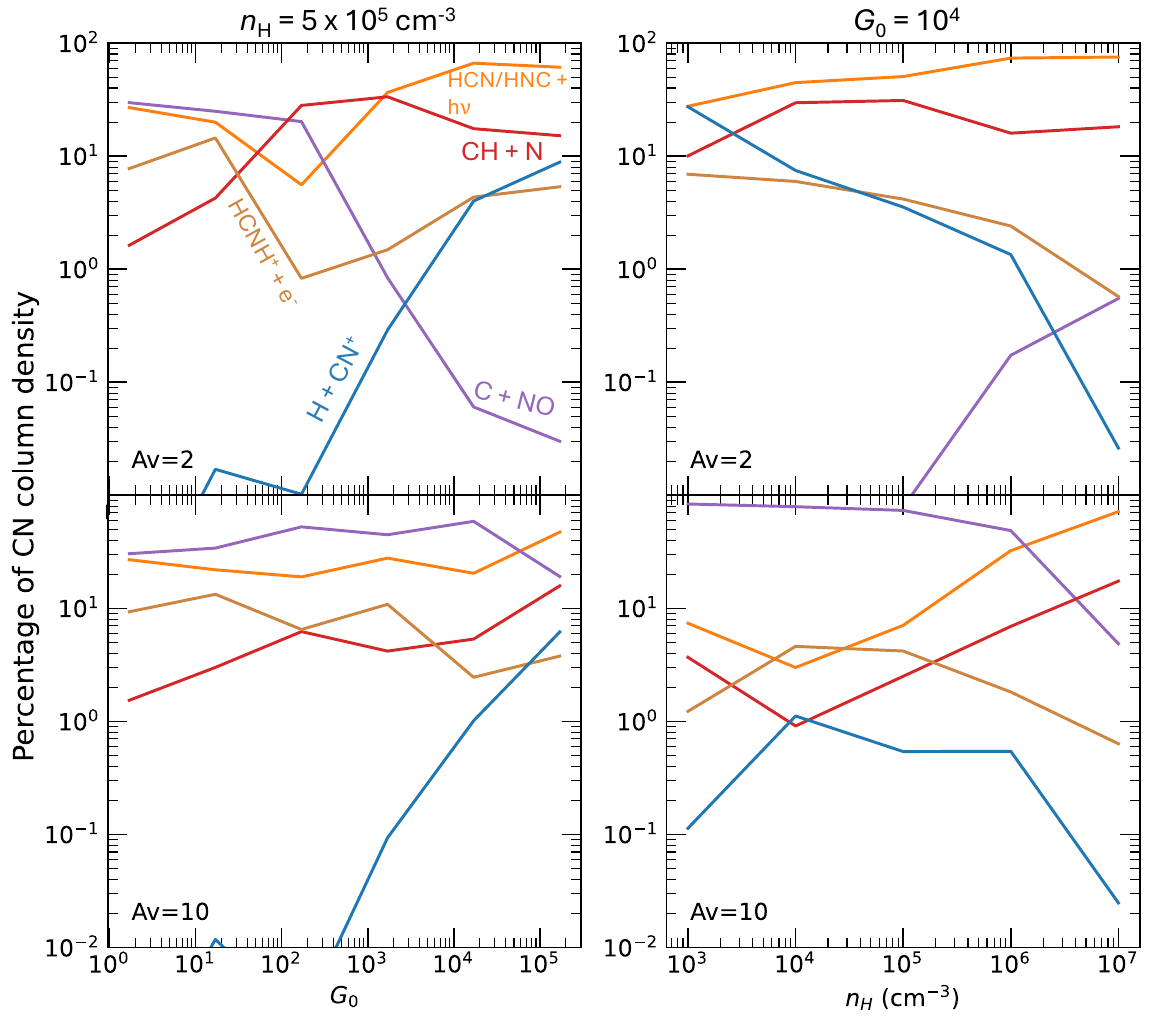}
    \caption{Importance of the formation pathways in the total column density of (top) HCN and (bottom) CN as a function of (left) $G_0$ and (right) gas density for different values of visual extinction $A_V$.}
    \label{fig:perc_form_coldens}
\end{figure}

To better estimate the importance of the warm UV-initiated chemistry locally,
we investigate its dependence on gas density and incident UV field. Hence, we run a grid of models with density ranging from $n_{\rm H} = 10^3-10^7$~cm$^{-3}$ and intensity of the UV field ranging from $G_0 = 1-10^5$. Figure \ref{fig:perc_form_coldens} shows the contribution of each formation pathway to the total column density, and reveals that the warm chemistry pathway (CN + H$_2$ $\rightarrow$ HCN + H) is highly dependent on the gas density and the intensity of the UV field (leading to high temperature and H$_2$ densities). Hence, at high density and intense UV field, the fraction of HCN formed by gas phase reaction with H$_2$, thus formed close to the H/H$_2$ transition, increases drastically. This result agrees with observations, where HCN peaks close to the H/H$_2$ transition. Indeed, the Orion Bar is a region with the necessary characteristics (high gas density $n_{\rm H} \sim 10^6$~cm$^{-3}$ and intense incident UV field $G_0 \sim 10^4$) to produce HCN through the warm chemistry. In addition, at high density and UV flux, the contribution of warm chemistry to the total HCN column density along the line of sight can become dominant (more than 30\% for $G_0>10^4$ and $n_{\rm H} = 10^6$~cm$^{-3}$). Hence, the emission from HCN produced in warm regions cannot be neglected, and HCN emission does not only trace denser, more shielded cold regions.

The bottom panel of Fig. \ref{fig:perc_form_coldens} shows that, contrary to HCN warm chemistry, the different formation pathways of CN, especially those linked to the UV irradiation (photodissociation of HCN/HNC and CH + N $\rightarrow$ CN + H), do not depend drastically on the physical conditions of the PDR. This result reveals that CN emission in the warm layers will be less affected by higher density or a stronger UV field than HCN emission. Hence, studying the variation of  HCN and CN column densities near the photodissociation front where the chemistry is dominated by the UV-initiated chemistry can be used to constrain the physical conditions. 

\subsubsection{Evidence of destruction of CN-functionalized PAHs?}
\label{sect:CN-PAHs}
Interestingly, we observe CN emission in the predominantly atomic region of the Orion Bar (see Fig. \ref{fig:MO_maps_ALMA}) at the systemic velocity of the Bar and not OMC-1 (see Fig. \ref{fig:velocity_maps}). In addition, the emission of CN at this position is very bright (only a factor 1.5 lower than the peak emission at the DFs) and the H$_2$/CN intensity ratio is too low in this region (even when correcting H$_2$ emission from extinction) for this structure to be a molecular filament. No background molecular structures are observed, only the foreground expanding shell of neutral atomic gas traced by C$^+$ \citep{Pabst20}. Hence, it is very unlikely that this CN emission originates from a different region than the Bar itself.  However, CN is not expected to be abundant in this region as it should be easily destroyed by UV photons, and the standard gas-phase chemistry predicted by 1D stationary models described in the previous section is not efficient to produce CN within these layers owing to the very low H$_2$ abundances (see Fig. \ref{fig:model_p1e8}). This predominatly atomic region, where molecules' abundances are very small, is also predicted by 1D dynamical models \citep[e.g.,][]{Bron_2018} and 3D PDR models \citep[][]{Bisbas_2012,Gaches_2026}. CN emission in the atomic region originates from the same region where faint C$_2$H emission is also observed, which may be the result of PAH or carbonaceous grain photoprocessing \citep{Goicoechea_2025}. As several CN-functionalized PAHs have been observed in the interstellar medium with significant abundances \citep[e.g.,][]{Mc_Guire_2018,McGuire_2021,Gabi_2024,Cernicharo_2024,Wenzel_2025} and it has been shown that CN radicals react readily with PAHs \citep[e.g.,][]{Heitkamper_2022}, the question is whether this emission could originate from the ongoing destruction of CN-functionalized PAHs. This hypothesis would be consistent with the possible top-down formation mechanisms discussed in \cite{Goicoechea_2025}, which could produce small hydrocarbons in predominantly atomic PDR environments. However, the complex 3D geometry and kinematics of the line of sight toward the Orion Bar makes it challenging to rationalize that CN emission in the atomic region is definitively linked to these chemical pathways and their analysis 
is beyond the scope of this paper.

\section{Conclusion}

In this work, we have studied rotationally excited emission of HCN and CN detected in the Orion Bar with ALMA at very high-angular resolution and established constraints on the gas density near the dissociation front, as well as on the formation and destruction pathways of these molecules, to investigate their morphology. The main conclusions of this study can be summarized as follows:

\begin{enumerate}
    \item HCN $4-3$ and CN $3-2$ emission have very similar morphology and peak very close to the DF, in disagreement with lower-angular-resolution observations which showed a stratified structure, where CN peaks closer to the UV-edge, while HCN peaks deeper in the UV-shielded gas. This is due to active warm chemistry driven by UV irradiation, involving reactions with warm or FUV-pumped H$_2$ and small hydrocarbons.
    \item HCN and CN emission is more extended toward the shielded gas than C$_2$H and H$_2$ emission, highlighting the importance of additional efficient formation pathways in colder gas.
    \item HCN $4-3$ emission presents two clear velocity components at $v_{\rm LSR} = 10.5$~km~s$^{-1}$ and  $v_{\rm LSR} = 11.5$~km~s$^{-1}$, which we find consistent with a scenario where the DF has been compressed by a UV-induced shock.
    \item The observations of multiple HCN and CN lines reveal a subthermal excitation $T_{\rm rot} \sim 10-30$~K, which implies that the gas density cannot be much higher than the critical densities of these lines.
    \item Both \texttt{RADEX} and Meudon PDR Code models point toward a density around $n_{\rm H} = 10^6$~cm$^{-3}$, hence a thermal pressure around $P_{\rm th}/k_B = 2 \times 10^8$~K~cm$^{-3}$ and suggest the existence of a pressure gradient at the DF.
    \item The moderately high density derived in all the field of view is in disagreement with the small-scale clumpiness hypothesis. The different tracers can, however, be explained by the observed density gradient.
    \item Detailed modeling of HCN and CN chemistry reveals the importance of warm chemistry enhanced by the UV field. Hence, in dense and irradiated environments, HCN can trace warm regions near the DF, not only the colder, denser gas traditionally associated with HCN emission.

\end{enumerate}

\begin{acknowledgements}
   
   This paper makes use of the ALMA data ADS/JAO.ALMA\#2015.1.01082.S. ALMA is a partnership of ESO (representing its member states), NSF (USA), and NINS (Japan), together with NRC (Canada), and NSC and ASIAA (Taiwan), in cooperation with the Republic of Chile. The Joint ALMA Observatory is operated by ESO, AUI/NRAO, and NAOJ. It also includes IRAM 30m telescope observations. IRAM is supported by INSU/CNRS (France), MPG (Germany), and IGN (Spain). M.Z., J.R.G. and S.C. thank the Spanish MCINN for funding support under grant PID2023-146667NB-I00. M. Z. acknowledges the Juan de la Cierva Postdoctoral Fellow project JDC2024-054658-I, funded by MICIU/AEI/10.13039/501100011033 and by the ESF+.
\end{acknowledgements}

\bibliographystyle{aa}
\bibliography{biblio}

\begin{appendix}\label{Sect:Appendix}

\section{Opacity of the lines}

\begin{figure}[!h]
    \centering
    \includegraphics[width=\linewidth]{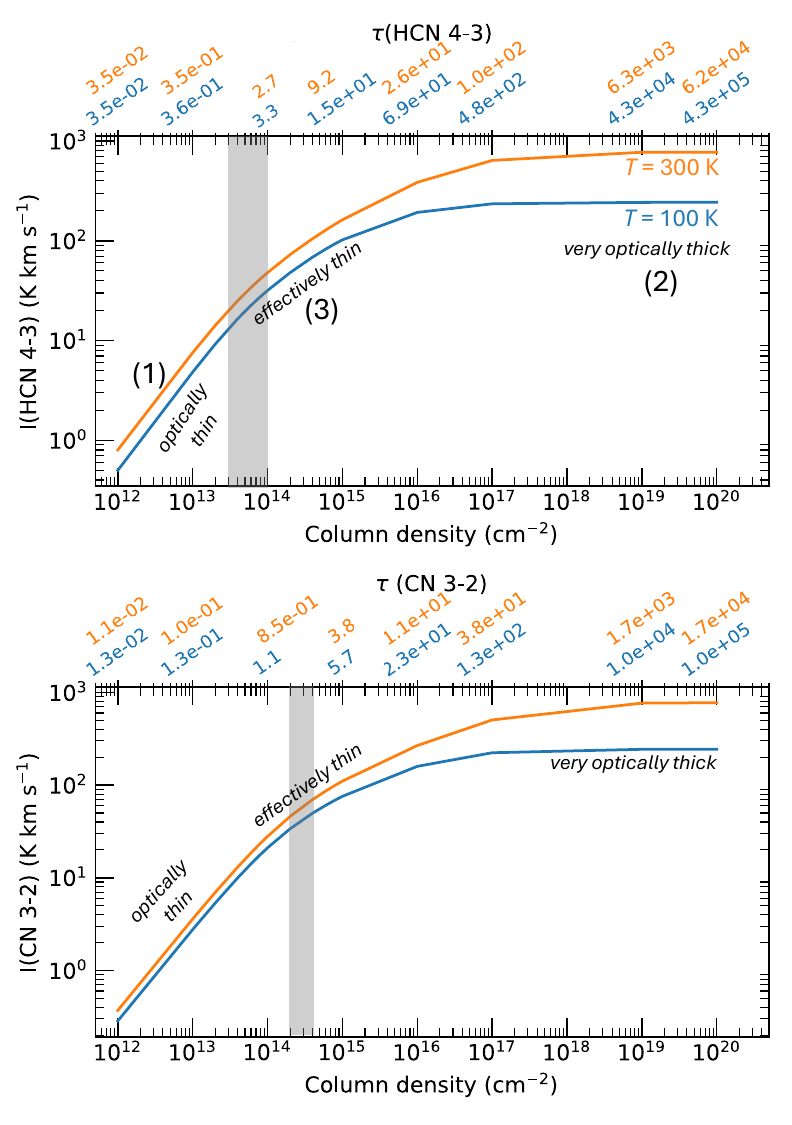}
    \caption{\texttt{RADEX} models showing the intensity of the HCN $4-3$  and CN $3-2$ lines as a function of their column density and line opacity for a gas density of $n_{\rm H} = 10^6$~cm$^{-3}$. The grayed area is the observed column density in the DF of the Orion Bar.}
    \label{fig:prop_I_Ncol}
\end{figure}

We use \texttt{RADEX} \citep{van_der_tak_2013} to compute the opacity at the center of the rotational lines, neglecting the fine and hyperfine components, of CN and HCN. We adopt the same line width as used in Sect. \ref{sect:radex}, ie, $\Delta v = 2.5$~km~s$^{-1}$. Fig. \ref{fig:prop_I_Ncol} displays the evolution of the optical thickness of HCN $4-3$ and CN $3-2$ lines as a function of column density. We can observe three different regimes. (1) At low column density ($N$(HCN) $<10^{13}$~cm$^{-2}$, $N$(CN) $<10^{14}$~cm$^{-2}$), the intensity of the lines is directly proportional to the column density: the line is optically thin. (2) At high column density ($N$(HCN) $>10^{15}$~cm$^{-2}$, $N$(CN) $>10^{16}$~cm$^{-2}$), the intensity saturates and does not depend on the column density anymore: the line is optically thick. (3) In between, the intensity is not exactly proportional to the column density but, due to the subthermal excitation ($T_{\rm rot} \ll T_{\rm gas}$), the deviation is very small, less than a factor of 2: the line is effectively thin. Assuming optically thin emission in this regime is thus a first-order approximation.

\label{opacity}

\section{Beam dilution: line intensity correction}\label{sec:correction-obs}

Because the beam size of the IRAM\,30m and \textit{Herschel}/HIFI  line surveys varies with frequency, the
observation of  \mbox{multiple rotational} lines of a given species (here HCN and CN) provides the line intensities averaged over slightly different areas of the Bar. In addition, these pointed observations do not spatially resolve the molecular emission arising from the PDR edge. To approximately  correct for these differences, we determined a frequency-dependent \mbox{\textit{beam coupling factor}} ($f_{\rm b}$) 
using the spatial information of the existing ALMA 
\mbox{HCN $J$\,=\,4--3} and \mbox{CN $N$\,=\,3--2} emission maps. In doing this, we
assume that all HCN  emission lines, and all CN
emission lines (including their isotopologues) have the same spatial distribution. 
Hence, we correct the single-dish line intensities as: 
\mbox{$W_{\rm corr,\,HCN}$\,=\,$W_{\rm obs,\,HCN} / f_{\rm b,\,HCN}$} and
\mbox{$W_{\rm corr,\,CN}$\,=\,$W_{\rm obs,\,CN} / f_{\rm b,\,CN}$}.
We do this in a two-step procedure.
We first smoothed the large  \mbox{HCN $J$\,=\,4--3} and \mbox{CN $N$\,=\,3--2} maps obtained with the IRAM\,30m telescope (\mbox{$\sim$8$''$ resolution}) to the different
HPBWs of each rotational line observed in the line surveys (\mbox{$J$\,=\,1--0} to \mbox{8--7} for HCN and 
\mbox{$N$\,=\,1--0} to \mbox{7--6} for CN;
providing HPBWs up to 28$''$ with the IRAM\,30m telescope and up to 40$''$ with Herschel). We then compute
 \mbox{$f_{\rm b}$(single-dish)\,=\,$W_{\rm smooth}$(HPBW)/$W_{\rm obs}$(8$''$)}, where
 \mbox{$W_{\rm smooth}$(HPBW)} is the  
 intensity (in K\,km\,s$^{-1}$) of the \mbox{HCN $J$\,=\,4--3} (or \mbox{CN $N$\,=\,3--2}) line extracted from the spatially
 smoothed maps toward the DF position. In a second step, we smooth the high angular resolution ALMA maps of these 
two lines to a 8$''$ angular resolution and compute
 \mbox{$f_{\rm b}$(ALMA)\,=\,$W_{\rm smooth}$(8$''$)/$W_{\rm obs}$(ALMA)}, where
 \mbox{$W_{\rm smooth}$(8$''$)} is the intensity
 of the  \mbox{HCN $J$\,=\,4--3} (or \mbox{CN $N$\,=\,3--2}) line 
  extracted from the smoothed ALMA maps toward the
 DF position.
  The final beam coupling factor, one for HCN lines and the other for CN lines, is 
 thus  \mbox{$f_{\rm b}$\,=\,$f_{\rm b}$(single-dish)\,$\cdot$$f_{\rm b}$(ALMA)}. These
  correction factors are listed in Table~\ref{table:f_b}.1. 
\begin{table*}[b]  
  \begin{center}
    \caption{Beam coupling correction factors obtained from IRAM\,30m and ALMA 
    HCN $J$\,=\,4$-$3 and CN $N$\,=\,3$-$2 maps. }
     \begin{tabular}{l   c c c c c  c@{\vrule height 10pt depth 5pt width 0pt}}   
      \hline \hline
      Species  /\,Receiver  & Frequency  &    Telescope      &     HPBW       &                     &               &      \\ 
       						& $[$GHz$]$  &   /\,Receiver     & $[$arcsec$]$   &  $f_b$(single-dish) &   $f_b$(ALMA) &  $f_b$(total)\\         
      \hline
HCN~$J=1-0$                 & 	88.631	 &  IRAM\,30m\,/\,E0 &      28        &  	  0.77		    &   0.93	    & 0.72  \\
HCN~$J=2-1$					&  177.261	 &	IRAM\,30m\,/\,E1 &      14        &		  0.86			&	0.93		& 0.80  \\
HCN~$J=3-2$	                &  265.886	 &	IRAM\,30m\,/\,E2 &       9        &		  0.90			&	0.93		& 0.84  \\
HCN~$J=4-3$					&  354.505	 &	IRAM\,30m\,/\,E3 &       8        &		  1.00			&	0.93		& 0.93  \\
HCN~$J=6-5$					&  531.716	 & Herschel\,/\,HIFI &      40        &		  0.69			&	0.93		& 0.65  \\
HCN~$J=7-6$					&  620.304	 & Herschel\,/\,HIFI &      34        &		  0.73			&	0.93		& 0.68  \\
HCN~$J=8-7$					&  708.877	 & Herschel\,/\,HIFI &      30        &		  0.75			&	0.93		& 0.71  \\
\hline
CN~$N=1-0$                  & $\sim$113  &  IRAM\,30m\,/\,E0 &      22        &  	  0.76  		&   0.89	    & 0.67  \\
CN~$N=2-1$					& $\sim$226  &	IRAM\,30m\,/\,E2 &      11        &		  0.90   	    &	0.89		& 0.80  \\
CN~$N=3-2$	                & $\sim$340  &	IRAM\,30m\,/\,E3 &       8        &		  1.00   		&	0.89		& 0.89  \\
CN~$N=5-4$                  & $\sim$567	 & Herschel\,/\,HIFI &      38        &  	  0.62 		    &   0.89		& 0.55  \\
CN~$N=6-5$					& $\sim$680  & Herschel\,/\,HIFI &      31        &		  0.67   		&	0.89		& 0.59  \\
CN~$N=7-6$	                & $\sim$793  & Herschel\,/\,HIFI &      27        &		  0.71   		&	0.89		& 0.63 \\\hline

      \end{tabular}
  \tablefoot{Rotational line intensities obtained
    with the IRAM\,30m and \textit{Herschel} telescopes toward the  DF position are corrected
    as \mbox{$W_{\rm corr}$\,=\,$W_{\rm obs}$\,/\,$f_b$(total)}, with \mbox{$f_b$(total)\,=\,$f_b$(single-dish)$\cdot$}$f_b$(ALMA). $W_{\rm obs}$(HIFI) are taken from Table A.1 of \cite{Nagy_2017}.}
  \end{center}
  \label{table:f_b}
\end{table*}

\onecolumn
\section{IRAM\,30m line observations of HCN and CN toward the dissociation front}

\begin{figure*}[th]
\includegraphics[width=\linewidth]{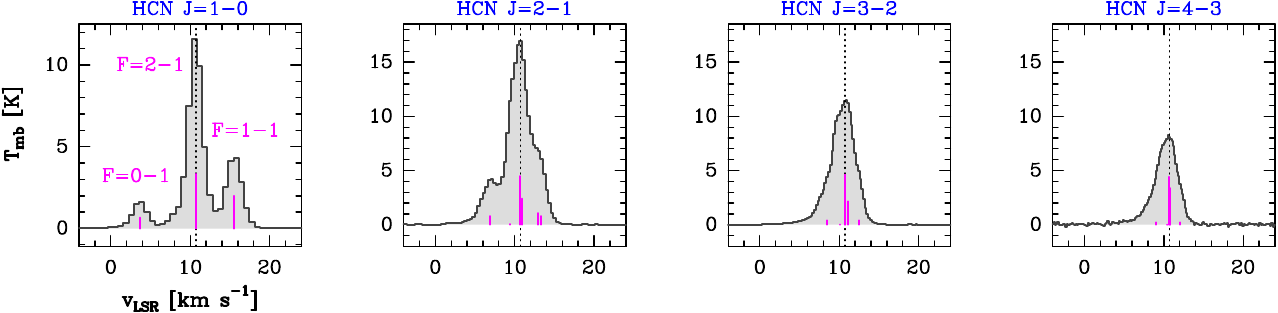}
\caption{IRAM\,30m observations of multiple rotational lines of HCN toward the Orion Bar DF position. Magenta lines show the position
of HFS lines and their relative strengths in the optically thin limit. } 
\label{fig:hcn_spectra30m}
\end{figure*}

\begin{figure*}[th]
\centering   
\includegraphics[width=\linewidth]{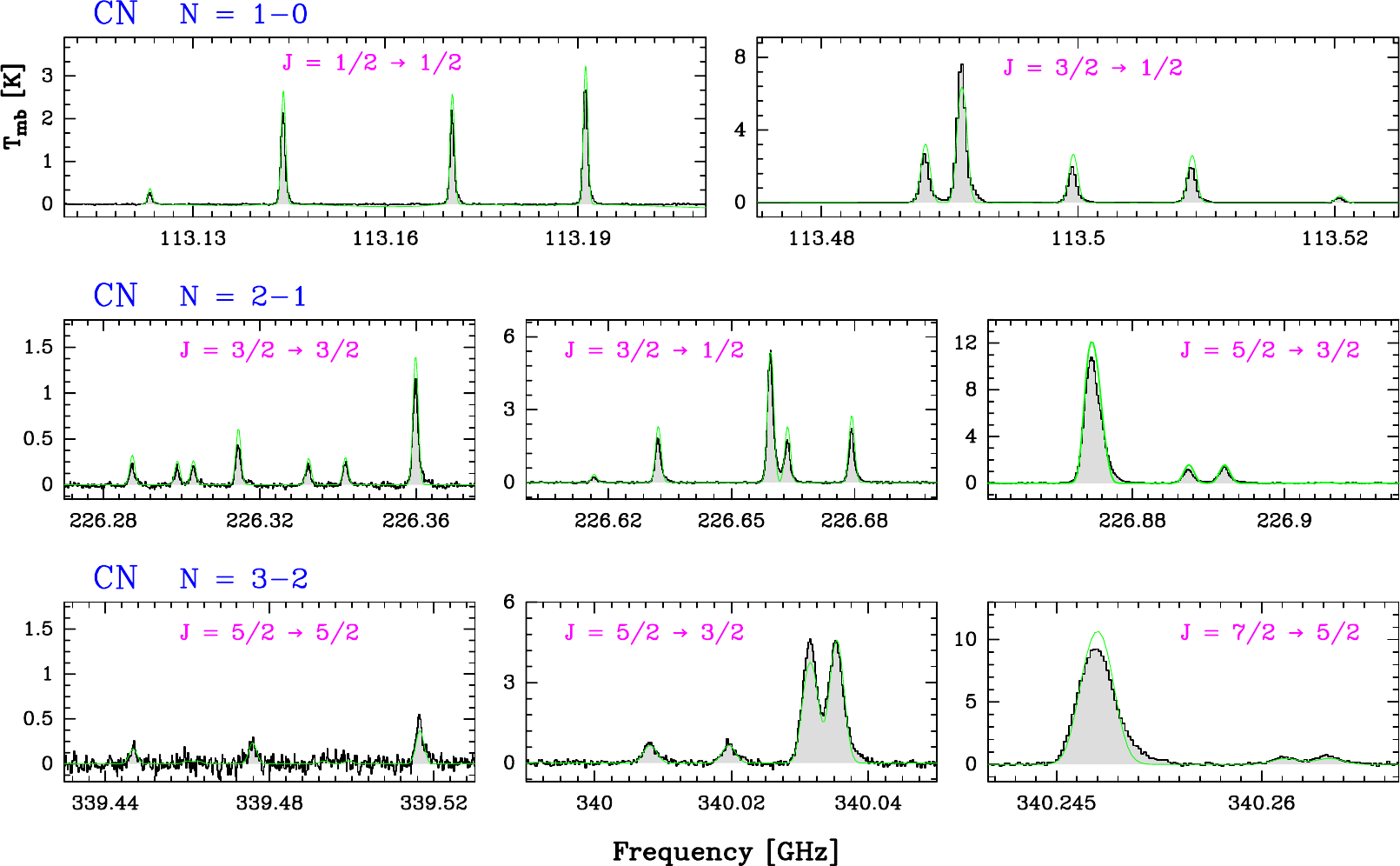}
\caption{IRAM\,30m observations of multiple rotational lines of CN toward the Orion Bar DF position. Each box shows different HFS lines. Green curves show a simple LTE fit
  with \mbox{$T_{\rm rot}$\,=\,12\,K} to better identify these lines.} 
\label{fig:cn_spectra30m}
\end{figure*}

       \begin{table*}[!h] 
       \begin{center}
       \caption{Line parameters of HCN.}  \label{Table_HCN}  
 \resizebox{1\textwidth}{!}{
  \begin{tabular}{l c c c r c c c c c c c c@{\vrule height 10pt depth 5pt width 0pt}}   
       \hline\hline       

  Species &   Transition &   Frequency &  HPBW$^{\#}$ & $E_{\rm u}$/$k$ & $A_{\rm ul}$ &  $S_{\rm ul}$ &  $g_{\rm u}$ &  $\displaystyle{\int} T_{\rm MB}$d$v$  &  $v_{\rm LSR}$ &  $\Delta v$ &  $T_{\rm MB}$ & $S/N$ \rule[-0.3cm]{0cm}{0.8cm}\ \\ \cline{2-2}

 & $J_{\rm u},F_{\rm u} \rightarrow J_{\rm l},F_{\rm l}$ & [MHz] & [$^{\prime\prime}$] & [K] & [s$^{-1}$] & & & [K~km~s$^{-1}$] & [km~s$^{-1}$] & [km~s$^{-1}$]  & [K] & \\

\hline

HCN$^{a}$    &   1,1 $\rightarrow$ 0,1 & \ \ 88630.415 & 28 & 4.3 &  2.41 $\times$ 10$^{-5}$ &  1.00 &   3  &   11.47\,(0.10)  &   10.7\,($<$0.1) &    2.4\,($<$0.1)  &   4.40   & 670  \\
             &   1,2 $\rightarrow$ 0,1 & \ \ 88631.847 & 28 & 4.3 &  2.41 $\times$ 10$^{-5}$ &  1.70 &   5  &   27.01\,(0.10)  &   10.7\,($<$0.1) &    2.2\,($<$0.1)  &  11.56   & 1850 \\
             &   1,0 $\rightarrow$ 0,1 & \ \ 88633.936 & 28 & 4.3 &  2.41 $\times$ 10$^{-5}$ &  0.33 &   1  &    3.95\,(0.10)  &   10.7\,($<$0.1) &    2.3\,($<$0.1)  &   1.61   & 250  \\

             &   2,2 $\rightarrow$ 1,2 & 177259.678 & 14  &  12.8 & 5.76 $\times$ 10$^{-5}$  & 0.50  &  5  &  \multirow{2}{*}[0cm]{$\Big \}$ 15.86\,(0.08)}  &   \multirow{2}{*}[-0.01cm]{10.7\,($<$0.1)} &   \multirow{2}{*}[-0.01cm]{1.9\,(0.1)}   &   \multirow{2}{*}[-0.01cm]{5.46}   & \multirow{2}{*}[-0.01cm]{\ \ 175} $\ddag$ \\             
             &   2,1 $\rightarrow$ 1,0 & 177259.924 & 14  &  12.8 & 1.28 $\times$ 10$^{-4}$  & 0.67  &  3  &               &                 &                 &      &   \\
             &   2,2 $\rightarrow$ 1,1 & 177261.111 & 14  &  12.8 & 1.73 $\times$ 10$^{-4}$  & 1.50  &  5  &               &                 &                 &      &   \\       
             &   2,3 $\rightarrow$ 1,2 & 177261.222 & 14  &  12.8 & 2.31 $\times$ 10$^{-4}$  & 2.80  &  7  &   \multirow{3}{*}[0.38cm]{$\Bigg \}$  41.62\,(0.14)}     &   10.7\,($<$0.1) &   2.6\,(0.1)   &  16.74  & \ \ 535 $\ddag$ \\  
             &   2,1 $\rightarrow$ 1,2 & 177262.012 & 14  &  12.8 & 6.42 $\times$ 10$^{-6}$  & 0.03  &  3  &               &                 &                 &      &   \\
             &   2,1 $\rightarrow$ 1,1 & 177263.445 & 14  &  12.8 & 9.63 $\times$ 10$^{-5}$  & 0.50  &  3  &   12.31\,(0.09)  &   10.8\,(0.1)    &   2.6\,(0.1)   &   3.89     & 125  \\      
             
             & \ \ 3 $\rightarrow$ 2$^{\dag}$     &  265886.433 & \ \  9 &  25.5 &  8.35 $\times$ 10$^{-4}$ &  3.00 &   21 &   42.59\,(0.05)  &   10.5\,(0.1) &    3.5\,(0.1)  &  11.27   & 930  \\
             & \ \ 4 $\rightarrow$ 3$^{\dag}$     &  354505.476 & \ \  7 &  42.5 &  2.05 $\times$ 10$^{-3}$ &  4.00 &   27 &   27.69\,(0.12)  &   10.4\,(0.1) &    3.2\,(0.1)  &   8.06   &  90  \\
             
             \hline                
     \end{tabular} 
     }
     \end{center}  
   \tablefoot{The spectroscopic line parameters were obtained from MADEX code \citep{Cerni_2012}. 
 Parentheses indicate the uncertainty obtained by the Gaussian fitting program. 
 $\#$~The half power beam width can be fitted by \mbox{HPBW[arcsec]\,$\approx$\,2460/Frequency[GHz]}.
   $\dag$~Hyperfine structure not resolved.
   $\ddag$~Fully blended transitions are marked with connecting symbols. Values obtained for the blended line of the overlapping hyperfine components.
  }
  \end{table*} 

\begin{table*}[!h] 
\begin{center}
\caption{Line parameters of CN.}  \label{Table_CN}  
\resizebox{1\textwidth}{!}{
\begin{tabular}{c c c c c c c c c c c c@{\vrule height 10pt depth 5pt width 0pt}}   
\hline\hline       

  Species, Transition &   Frequency &  HPBW & $E_{\rm u}$/$k$ & $A_{\rm ul}$ &  $S_{\rm ul}$ &  $g_{\rm u}$ &  $\displaystyle{\int} T_{\rm MB}$d$v$  &  $v_{\rm LSR}$ &  $\Delta v$ &  $T_{\rm MB}$ & $S/N$ \rule[-0.3cm]{0cm}{0.8cm}\ \\

  & [MHz] & [$^{\prime\prime}$] & [K] & [s$^{-1}$] & & & [K~km~s$^{-1}$] & [km~s$^{-1}$] & [km~s$^{-1}$]  & [K] & \\

  \hline                                           
   {\bf CN}, $(N, J, F)_{\rm u} \rightarrow (N, J, F)_{\rm l}$  &  &  &   &   &   &    &     &    &    &   & \\ \cline{1-1}

   1,1/2,1/2 $\rightarrow$ 0,1/2,1/2    &  113123.370   & 22     & \ \ 5.4   &  1.29 $\times$ 10$^{-6}$   & 0.07   &  2    &  0.58\,(0.01)  &   10.8\,($<$0.1)   &   2.1\,($<$0.1)   &  0.26   & 38  \\  
   1,1/2,1/2 $\rightarrow$ 0,1/2,3/2    &  113144.157   & 22     & \ \ 5.4   &  1.05 $\times$ 10$^{-5}$   & 0.59   &  2    &  4.58\,(0.02)  &   10.9\,($<$0.1)   &   2.1\,($<$0.1)   &  2.08   & 280 \\  
   1,1/2,3/2 $\rightarrow$ 0,1/2,1/2    &  113170.492   & 22     & \ \ 5.4   &  5.15 $\times$ 10$^{-6}$   & 0.58   &  4    &  4.69\,(0.02)  &   10.8\,($<$0.1)   &   2.1\,($<$0.1)   &  2.12   & 280  \\ 
   1,1/2,3/2 $\rightarrow$ 0,1/2,3/2    &  113191.279   & 22     & \ \ 5.4   &  6.68 $\times$ 10$^{-6}$   & 0.75   &  4    &  5.95\,(0.03)  &   10.8\,($<$0.1)   &   2.1\,($<$0.1)   &  2.71   & 350 \\ 
   1,3/2,3/2 $\rightarrow$ 0,1/2,1/2    &  113488.120   & 22     & \ \ 5.4   &  6.74 $\times$ 10$^{-6}$   & 0.75   &  4    &  5.61\,(0.11)  &   10.9\,($<$0.1)   &   2.0\,($<$0.1)   &  2.57   & 360 \\ 
   1,3/2,5/2 $\rightarrow$ 0,1/2,3/2    &  113490.970   & 22     & \ \ 5.4   &  1.19 $\times$ 10$^{-5}$   & 2.00   &  6    & 16.68\,(0.08)  &   10.9\,($<$0.1)   &   2.1\,($<$0.1)   &  7.55   & 1000 \\
   1,3/2,1/2 $\rightarrow$ 0,1/2,1/2    &  113499.644   & 22     & \ \ 5.4   &  1.06 $\times$ 10$^{-5}$   & 0.59   &  2    &  4.13\,(0.03)  &   10.9\,($<$0.1)   &   2.1\,($<$0.1)   &  1.89   & 260 \\ 
   1,3/2,3/2 $\rightarrow$ 0,1/2,3/2    &  113508.907   & 22     & \ \ 5.4   &  5.19 $\times$ 10$^{-6}$   & 0.58   &  4    &  4.27\,(0.03)  &   10.8\,($<$0.1)   &   2.1\,($<$0.1)   &  1.94   & 260 \\ 
   1,3/2,1/2 $\rightarrow$ 0,1/2,3/2    &  113520.432   & 22     & \ \ 5.4   &  1.30 $\times$ 10$^{-6}$   & 0.07   &  2    &  0.49\,(0.01)  &   10.9\,($<$0.1)   &   1.9\,($<$0.1)   &  0.24   & 30  \\
   2,3/2,1/2 $\rightarrow$ 1,3/2,1/2    &  226287.419   & 11     &     16.3  &  1.03 $\times$ 10$^{-5}$   & 0.07   &  2    &  0.49\,(0.02)  &   10.8\,($<$0.1)   &   2.2\,(0.1)      &  0.21   & 15 \\  
   2,3/2,1/2 $\rightarrow$ 1,3/2,3/2    &  226298.943   & 11     &     16.3  &  8.23 $\times$ 10$^{-6}$   & 0.06   &  2    &  0.35\,(0.02)  &   10.8\,($<$0.1)   &   1.9\,(0.1)      &  0.18   & 15 \\  
   2,3/2,3/2 $\rightarrow$ 1,3/2,1/2    &  226303.037   & 11     &     16.3  &  4.17 $\times$ 10$^{-6}$   & 0.06   &  4    &  0.42\,(0.02)  &   10.8\,($<$0.1)   &   2.1\,(0.1)      &  0.19   & 15 \\ 
   2,3/2,3/2 $\rightarrow$ 1,3/2,3/2    &  226314.540   & 11     &     16.3  &  9.90 $\times$ 10$^{-6}$   & 0.14   &  4    &  0.85\,(0.02)  &   10.8\,($<$0.1)   &   1.9\,(0.1)      &  0.41   & 25 \\ 
   2,3/2,3/2 $\rightarrow$ 1,3/2,5/2    &  226332.499   & 11     &     16.3  &  4.56 $\times$ 10$^{-6}$   & 0.06   &  4    &  0.44\,(0.02)  &   10.9\,($<$0.1)   &   2.1\,(0.1)      &  0.20   & 15 \\ 
   2,3/2,5/2 $\rightarrow$ 1,3/2,3/2    &  226341.930   & 11     &     16.3  &  3.16 $\times$ 10$^{-6}$   & 0.07   &  6    &  0.49\,(0.02)  &   10.7\,($<$0.1)   &   2.0\,(0.1)      &  0.24   & 15 \\
   2,3/2,5/2 $\rightarrow$ 1,3/2,5/2    &  226359.871   & 11     &     16.3  &  1.61 $\times$ 10$^{-5}$   & 0.34   &  6    &  2.41\,(0.03)  &   10.8\,($<$0.1)   &   2.1\,($<$0.1)   &  1.08   & 65 \\ 
   2,3/2,1/2 $\rightarrow$ 1,1/2,3/2    &  226616.571   & 11     &     16.3  &  1.07 $\times$ 10$^{-5}$   & 0.08   &  2    &  0.45\,(0.02)  &   10.7\,($<$0.1)   &   2.1\,(0.1)      &  0.21   & 15 \\ 
   2,3/2,3/2 $\rightarrow$ 1,1/2,3/2    &  226632.190   & 11     &     16.3  &  4.26 $\times$ 10$^{-5}$   & 0.60   &  4    &  3.87\,(0.03)  &   10.8\,($<$0.1)   &   2.1\,($<$0.1)   &  1.72   & 105  \\
   2,3/2,5/2 $\rightarrow$ 1,1/2,3/2    &  226659.558   & 11     &     16.3  &  9.47 $\times$ 10$^{-5}$   & 1.99   &  6    & 12.00\,(0.02)  &   10.8\,($<$0.1)   &   2.2\,($<$0.1)   &  5.13   & 310  \\  
   2,3/2,1/2 $\rightarrow$ 1,1/2,1/2    &  226663.693   & 11     &     16.3  &  8.47 $\times$ 10$^{-5}$   & 0.59   &  2    &  4.01\,(0.06)  &   10.7\,($<$0.1)   &   2.5\,($<$0.1)   &  1.50   & 100  \\  
   2,3/2,3/2 $\rightarrow$ 1,1/2,1/2    &  226679.311   & 11     &     16.3  &  5.27 $\times$ 10$^{-5}$   & 0.74   &  4    &  4.58\,(0.03)  &   10.8\,($<$0.1)   &   2.1\,($<$0.1)   &  2.05   & 130  \\ 
   2,5/2,5/2 $\rightarrow$ 1,3/2,3/2    &  226874.191   & 11     &     16.3  &  9.62 $\times$ 10$^{-5}$   & 2.02   &  6    &              &                  &                 &         &     \\ 
   2,5/2,7/2 $\rightarrow$ 1,3/2,5/2    &  226874.781   & 11     &     16.3  &  1.14 $\times$ 10$^{-4}$   & 3.20   &  8    &  \multirow{3}{*}[0.38cm]{$\Bigg \}$ 36.53\,(0.04)}  & 10.8\,($<$0.1)  &  3.3\,($<$0.1)  &  10.34 & 605  \\ 
   2,5/2,3/2 $\rightarrow$ 1,3/2,1/2    &  226875.896   & 11     &     16.3  &  8.59 $\times$ 10$^{-5}$   & 1.20   &  4    &              &                  &                 &         &    \\
   2,5/2,3/2 $\rightarrow$ 1,3/2,3/2    &  226887.420   & 11     &     16.3  &  2.73 $\times$ 10$^{-5}$   & 0.38   &  4    &  2.41\,(0.03)  &   10.8\,($<$0.1)   &   2.1\,($<$0.1)   &  1.10   & 70   \\ 
   2,5/2,5/2 $\rightarrow$ 1,3/2,5/2    &  226892.128   & 11     &     16.3  &  1.81 $\times$ 10$^{-5}$   & 0.38   &  6    &  2.95\,(0.04)  &   10.8\,($<$0.1)   &   2.2\,($<$0.1)   &  1.26   & 80   \\ 
   2,5/2,3/2 $\rightarrow$ 1,3/2,5/2    &  226905.357   & 11     &     16.3  &  1.13 $\times$ 10$^{-6}$   & 0.02   &  4    &  0.09\,(0.01)  &   10.6\,(0.1)      &   1.5\,(0.3)      &  0.06   & 4    \\
   3,5/2,3/2 $\rightarrow$ 2,5/2,3/2    &  339446.777   & \ \ 7  &     32.6  &  2.26 $\times$ 10$^{-5}$   & 0.09   &  4    &  0.32\,(0.05)  &   10.7\,(0.1)      &   1.6\,(0.3)      &  0.19   & 6  \\  
   3,5/2,5/2 $\rightarrow$ 2,5/2,5/2    &  339475.904   & \ \ 7  &     32.6  &  2.12 $\times$ 10$^{-5}$   & 0.13   &  6    &  0.48\,(0.06)  &   10.8\,(0.1)      &   2.1\,(0.3)      &  0.21   & 7   \\ 
   3,5/2,7/2 $\rightarrow$ 2,5/2,7/2    &  339516.635   & \ \ 7  &     32.6  &  2.54 $\times$ 10$^{-5}$   & 0.21   &  8    &  0.90\,(0.05)  &   10.8\,($<$0.1)   &   1.8\,(0.1)      &  0.47   & 10  \\ 
   3,5/2,5/2 $\rightarrow$ 2,3/2,5/2    &  340008.126   & \ \ 7  &     32.6  &  6.20 $\times$ 10$^{-5}$   & 0.39   &  6    &  1.44\,(0.06)  &   10.7\,($<$0.1)   &   2.0\,(0.1)      &  0.69   & 15  \\ 
   3,5/2,3/2 $\rightarrow$ 2,3/2,3/2    &  340019.626   & \ \ 7  &     32.6  &  9.27 $\times$ 10$^{-5}$   & 0.39   &  4    &  1.49\,(0.08)  &   10.6\,($<$0.1)   &   2.1\,(0.1)      &  0.67   & 15   \\
   3,5/2,7/2 $\rightarrow$ 2,3/2,5/2    &  340031.549   & \ \ 7  &     32.6  &  3.85 $\times$ 10$^{-4}$   & 3.20   &  8    & 10.04\,(0.10)  &   10.8\,($<$0.1)   &   2.1\,($<$0.1)   &  4.54   & 100   \\ 
   3,5/2,3/2 $\rightarrow$ 2,3/2,1/2    &  340035.408   & \ \ 7  &     32.6  &  2.89 $\times$ 10$^{-4}$   & 1.20   &  4    & \multirow{2}{*}[0cm]{$\Big \}$ 11.19\,(0.11)}  &   \multirow{2}{*}[-0.01cm]{10.8\,($<$0.1)}   &   \multirow{2}{*}[-0.01cm]{2.4\,($<$0.1)}   & \multirow{2}{*}[-0.01cm]{4.36}   & \multirow{2}{*}[-0.01cm]{100}   \\ 
   3,5/2,5/2 $\rightarrow$ 2,3/2,3/2    &  340035.408   & \ \ 7  &     32.6  &  3.23 $\times$ 10$^{-4}$   & 2.01   &  6    &              &                  &                 &         &      \\
   3,7/2,7/2 $\rightarrow$ 2,5/2,5/2    &  340247.770   & \ \ 7  &     32.7  &  3.80 $\times$ 10$^{-4}$   & 3.15   &  8    &              &                  &                 &         &      \\ 
   3,7/2,9/2 $\rightarrow$ 2,5/2,7/2    &  340247.770   & \ \ 7  &     32.7  &  4.13 $\times$ 10$^{-4}$   & 4.29   & 10    &  \multirow{3}{*}[0.38cm]{$\Bigg \}$ 27.38\,(0.09)}  &   10.7\,($<$0.1)   &   2.8\,($<$0.1)   &  9.23   & 200  \\
   3,7/2,5/2 $\rightarrow$ 2,5/2,3/2    &  340248.544   & \ \ 7  &     32.7  &  3.67 $\times$ 10$^{-4}$   & 2.29   &  6    &              &                  &                 &         &      \\ 
   3,7/2,5/2 $\rightarrow$ 2,5/2,5/2    &  340261.773   & \ \ 7  &     32.7  &  4.48 $\times$ 10$^{-5}$   & 0.28   &  6    &  1.00\,(0.08)  &   10.8\,($<$0.1)   &   1.6\,(0.1)      &  0.57   & 15   \\
   3,7/2,7/2 $\rightarrow$ 2,5/2,7/2    &  340264.949   & \ \ 7  &     32.7  &  3.35 $\times$ 10$^{-5}$   & 0.28   &  8    &  1.55\,(0.09)  &   10.8\,($<$0.1)   &   2.2\,(0.2)      &  0.66   & 20   \\ 
    \hline   
   \end{tabular}
   }
\end{center} 
   \tablefoot{Frequencies, $E_u$, $A_{ul}$, $S_{\rm ul}$, and $g_u$ from CDMS catalogue \citep{Endres_2016}.}
   \end{table*}

\end{appendix}

\end{document}